\documentclass[useAMS,usenatbib]{mn2e}
\PassOptionsToPackage{hyphens}{url}\usepackage{hyperref}
\usepackage{epsfig,lscape} 
\usepackage{hyperref} 
\usepackage{natbib}
\usepackage[caption=false]{subfig}
\usepackage{lscape}
\usepackage{textcomp}
\usepackage{gensymb}
\usepackage{xcolor}
\usepackage{amsmath}
\usepackage{comment}
\usepackage{arydshln}
\usepackage{graphics}
\usepackage{booktabs}
\usepackage[caption=false]{subfig}
\usepackage{amsmath,amssymb}
\usepackage[normalem]{ulem}

\newcommand{\mnras}{MNRAS}
\newcommand{\apj}{ApJ}
\newcommand{\araa}{ARA\&A}
\newcommand{\aap}{A\&A}
\newcommand{\aj}{AJ}
\newcommand{\apjl}{ApJL}
\newcommand{\apjs}{ApJS}
\newcommand{\pasp}{PASP}
\newcommand{\ao}{ApOpt}
\newcommand{\nat}{Nature}

\title[Density distribution and structure of MAPs]{The Milky Way tomography with APOGEE: intrinsic density distribution and structure of mono-abundance populations}

\author[J. Lian et al.]{{Jianhui~Lian}$^{1,2}$\thanks{lian@mpia.de, gail.zasowski@gmail.com}, {Gail Zasowski}$^{2}$, {Ted Mackereth}$^{3}$, Julie Imig$^{4}$, Jon A. Holtzman$^{4}$,  
\newauthor Rachael L. Beaton$^{5,6}$, Jonathan C. Bird$^{7}$, Katia Cunha$^{8,9}$, Jos\'e G. Fern\'andez-Trincado$^{10}$, 
\newauthor Danny Horta$^{11}$, 
 Richard R. Lane$^{12}$, Karen Masters$^{13}$, 
 Christian Nitschelm$^{14}$, 
\newauthor {A. Roman-Lopes}$^{15}$\\
\small $^{1}${Max Planck Institute for Astronomy, 69117, Heidelberg, Germany}\\
\small $^{2}${Department of Physics \& Astronomy, University of Utah, Salt Lake City, UT 84112, USA}\\
\small $^{3}${Dunlap Institute for Astronomy and Astrophysics, University of Toronto, Toronto, Canada}\\
\small $^{4}${Department of Astronomy, New Mexico State University, Las Cruces, NM 88003, USA}\\
\small $^5${The Observatories of the Carnegie Institution for Science, 813 Santa Barbara Street, Pasadena, CA 91101, USA}\\
\small $^6${Department of Astrophysical Sciences, Princeton University, 4 Ivy Lane, Princeton, NJ, 08544, USA}\\
\small $^{7}${Department of Physics and Astronomy, Vanderbilt University, 6301 Stevenson Center, Nashville, TN 37235, USA}\\
\small $^{8}$ {Steward Observatory, The University of Arizona, Tucson, AZ, 85719, USA} \\
\small $^{9}$ {Observat\'orio Nacional, 20921-400 So Crist\'ovao, Rio de Janeiro, RJ, Brazil}\\
\small $^{10}$ {Instituto de Astronom\'ia, Universidad Cat\'olica
del Norte, Av. Angamos 0610, Antofagasta, Chile}\\
\small $^{11}${Astrophysics Research Institute, Brownlow Hill, Liverpool, L3 5RF, UK} \\
\small $^{12}${Instituto de Astrofısica, Pontificia Universidad Cat\'olica de Chile, Av. Vicuna Mackenna 4860, 782-0436 Macul, Santiago, Chile}\\
\small $^{13}${Departments of Physics and Astronomy, Haverford College, 370 Lancaster Avenue, Haverford, PA 19041, USA}\\
\small $^{14}${Centro de Astronom\'ia (CITEVA), Universidad de Antofagasta, Avenida Angamos 601, Antofagasta 1270300, Chile}\\
\small $^{15}${Departamento de Astronom\'ia, Universidad La Serena, La Serena, Chile}\\
}

\begin{document}

\maketitle

\begin{abstract}
The spatial distribution of mono-abundance populations (MAPs, selected in [Fe/H] and [Mg/Fe]) reflect the chemical and structural evolution in a galaxy and impose strong constraints on galaxy formation models. In this paper, we use APOGEE data to derive the intrinsic density distribution of MAPs in the Milky Way, after carefully considering the survey selection function. 
We find that a single exponential profile is not a sufficient description of the Milky Way's disc. Both the individual MAPs and the integrated disc exhibit a broken radial density distribution; densities are relatively constant with radius in the inner Galaxy and rapidly decrease beyond the break radius. 
We fit the intrinsic density distribution as a function of radius and vertical height with a 2D density model that considers both a broken radial profile and radial variation of scale height {(i.e., flaring)}. 
There is a large variety of structural parameters between different MAPs, indicative of strong structure evolution of the Milky Way.   
One surprising result is that high-$\alpha$ MAPs show the strongest flaring. 
The young, solar-abundance MAPs present the shortest scale height and least flaring, suggesting recent and ongoing star formation confined to the disc plane. Finally we derive the intrinsic density distribution and corresponding structural parameters of the chemically defined thin and thick discs. 
The {chemical} thick and thin discs have local surface mass densities of 5.62$\pm$0.08 and 15.69$\pm$0.32 
${\rm M_{\odot}pc^{-2}}$, 
respectively, suggesting a massive thick disc with a local surface mass density ratio between thick to thin disc of 36\%.  

\end{abstract}

\section{Introduction}
The stellar structure of the Milky Way, along with the chemical and kinematic configurations, place critical constraints on models of our Galaxy's formation and evolution. 
Our position at the Solar circle with a small vertical distance from the disc plane allows us to study the Galactic disc structure in great detail on a star by star basis. The disc 
of the Milky Way is generally thin, providing us a relatively unobscured view in the vertical direction into the stellar halo but highly obscured view in the disc plane, especially at the Galactic center direction. As a consequence, the local surface density and vertical density profile have been robustly measured \citep[e.g.,][]{gilmore1983,bienayme1987,robin2003,flynn2006,juric2008,mckee2015}, while the radial and vertical structure beyond the solar radius are more uncertain. 

An exponential form is commonly used to describe the radial and vertical density distribution of the Galactic and extragalactic disc \citep[e.g.,][]{freeman1970,gilmore1983,robin2003,pohlen2006}. 
Early studies of the Galaxy structure is generally based on stellar photometric observations without distance information \citep[e.g.,][]{bahcall1980}. The derived structure parameters by fitting simple models to the projected star counts usually suffer large uncertainties due to degeneracy between different parameters. While the scale length of external galaxies are well established \citep[e.g.,][]{fathi2010,lang2015}, a wide range of values from 1.8 to 6.0~kpc have been reported for the Milky Way's disc in the literature (see \citealt{bland2016} for a review). Comparing to external galaxies at similar stellar masses, the scale length of Milky Way's disc seems systematically shorter \citep{licquia2016,boardman2020b}. This suggests the Milky Way might be an unusually compact galaxy for its mass, or there are 
inconsistences between the methods used for scale length measurements in the Milky Way and other galaxies.   

Star counts in the vertical direction in early studies revealed that the Milky Way's disc consists of two components with distinct thickness, which are commonly referred to as the {\sl geometric} `thick' and `thin' discs \citep{yoshii1982,gilmore1983,robin2003,juric2008}. 
After being discovered in external galaxies by \citet{burstein1979}, 
such geometric thick/thin disc dichotomy was found to be common in local disc galaxies \citep{yoachim2006,comeron2012}. 
\citet{gilmore1983} provided the first reliable vertical stellar density distribution and estimated a scale height ($h_{\rm Z}$) of $1350$~pc for the thick disc and 300~pc for the thin disc. Similar values are suggested in \citet{siegel2002} using improved photometric data from wide-field CCDs. 
With a large sample of M dwarfs in SDSS photometry survey, \citet{juric2008} obtained a $h_{\rm Z}$ of 300~pc for the thin disc in good agreement with previous works, and 900~pc for the thick disc lower than earlier estimates. Unlike the scale height at solar radius, the scale length ($h_{\rm R}$) is more difficult to measure due to the high extinction on the disc plane. \citet{juric2008} reported a scale length of the thin disc shorter than the thick disc (2.9 versus 3.6~kpc). However, an opposite result is found in \citet{cheng2012} with a scale length of 3.4~kpc for the thin disc and 1.8~kpc for the thick disc. Interestingly, deep imaging data reveals comparable scale length between the thick and thin disc in external galaxies \citep{comeron2012}.  

In addition to the geometric dichotomy, many works have also identified a dichotomy in chemical compositions of stars in the solar neighborhood with two well separated sequences in [$\alpha$/Fe]-[Fe/H] distribution \citep[e.g.,][]{fuhrmann1998,reddy2006,lee2011,adibekyan2012,haywood2013,bensby2014,hayden2015}. This geometric and chemical dichotomy in disc stars are not identical, but they share significant overlap.   
While some chemical thin disc stars may be found in the geometric thick disc, it is characterized by old, $\alpha$-enhanced, and kinematically hot populations, while the geometric thin disc mostly consist of younger, solar-$\alpha$, and kinematically cooler populations \citep[e.g.,][]{bensby2005,lian2020b}. 
This connection between the geometric and chemical dichotomy implies close relation between the thick/thin disc structure formation and the chemical evolution history 
of the Galaxy \citep{lian2020b,horta2021}. 

A variety of scenarios have been suggested to explain the chemical dichotomy and its implication for thick/thin disc formation. One of them attributes the formation of metal-poor, low-$\alpha$ populations to a recent gas accretion and star burst event \citep{calura2009,haywood2019,spitoni2019,buck2020,lian2020a,lian2020b,agertz2021}. In this scenario, the thick and thin discs are established locally and successively. In contrast, another scenario assumes the high- and low-$\alpha$ sequences formed in parallel but at different locations in the Galaxy and mixed up later on via radial migration \citep{grisoni2017,mackereth2018}. The mixing through radial migration is also required in another explanation that adopts a continuously varying star formation history with radius \citep{chiappini2009,minchev2015,andrews2017,sharma2020,johnson2021}. In addition, clumpy star formation is also suggested to be capable of generating the chemical dichotomy \citep{clarke2019}.   

Recently, with chemical observations being available beyond the solar neighborhood, many works have confirmed the nearly invariant locus of high-$\alpha$ sequence, which is usually referred to as `chemical thick disc', in [$\alpha$/Fe]-[Fe/H] distribution across the Galaxy \citep[e.g.,][]{weinberg2019,katz2021}. This suggests highly homogeneous chemical enrichment in the early times or thoroughly mixing after the establishment of high-$\alpha$ disc.  
The characteristics of geometric thick disc, however, changes from the inner to the outer disc, where the region high above the disc is increasingly dominated by intermediate age, metal-poor, low-$\alpha$ populations \citep{nidever2014,hayden2015,queiroz2020}. This results in a radial age gradient in the geometric thick disc \citep{martig2016-age-gradient}. This finding complicates the formation picture of Milky Way's disc and illustrate the importance of studying the Galaxy beyond the solar vicinity to draw a comprehensive picture of the formation and evolution of the Milky Way.    

The advent of massive stellar spectrosopic surveys, which cover a large portion of the Galaxy, such as APOGEE \citep{majewski2017}, GALAH \citep{martell2016}, LAMOST \citep{zhao2012}, and Gaia-ESO \citep{gilmore2012}, have enabled the studies of spatial and chemical structure beyond the solar radius. Taking advantage of the wide spatial coverage of these surveys, many works have explored the structure of mono-abundance/age populations which reflects the growth history of the Milky Way and provide critical insights into mechanisms that drive the disc formation and evolution. 
\citet{bovy2012b} studied the mono-abundance populations (MAPs; defined in [Fe/H] and [$\alpha$/Fe] space) in the Milky Way using a sample of G-dwarfs in SEGUE survey. They found a radially compact, but vertically extended morphology of high-$\alpha$ disc, and a continuous transformation from the thick high-$\alpha$ disc to the thin low-$\alpha$ disc in spite of discontinuity in chemical abundance distribution. 
\citet{bovy2016} extended this work to red-clump giants in APOGEE survey and suggested a broken profile is needed to well fit the radial density distribution of the low-$\alpha$ disc. 
The density model is also improved on the basis of \citet{bovy2012b} to consider possible radial variation of scale height for each MAP. The authors found a constant thickness of high-$\alpha$ MAPs, but the scale height of low-$\alpha$ MAPs increases with radius (i.e. flares).   
This broken radial profile and flaring of low-$\alpha$ disc is confirmed in \citet{mackereth2017} where the authors dissect the disc stars for the first time into mono-age, mono-abundance populations. 
More recently, \citet{yu2021} studied the structure of MAPs using LAMOST data, and found a clear signature of flaring in both high- and low-$\alpha$ MAPs. 
Many theoretical works have also paid attention to the mono-abundance/age populations in simulated galaxies to understand the observed structure of MAPs in the Milky Way \citep[e.g.,][]{stinson2013,bird2013,minchev2015}. Although facing many challenges to match the high-dimensional observations in the Milky Way, the mono-abundance/age populations of simulated galaxies have structure in qualitative agreement with the Milky Way \citep[e.g.,][]{stinson2013,bird2013,martig2014a}.  

In this paper we investigate the structure of MAPs in the Milky Way using the latest APOGEE observations which incorporates the data recently acquired at Las Campanas Observatory at south hemisphere and provide a better coverage of the inner Galaxy. Most previous MAP studies have used a forward modelling approach which first predicts the number of stars at each Galactic location probed by the survey by applying the survey selection function to a presumed Milky Way density model and then tune the model parameters to achieve a good match to the data \citep{bovy2012b,bovy2016,mackereth2017}. With this approach, the intrinsic density distribution of MAPs are unknown until a good fit by the model is achieved.  
In this work we adopt a different method to first recover the spatial density distribution of the underlying {\sl intrinsic} population, including stars not observed, for each MAP from observations after carefully accounting for the survey selection function. We then fit a density model to the intrinsic, selection-function corrected, density distribution to derive structural parameters of MAPs. 
In this way we can construct a more realistic and representative density model with the guidance from the intrinsic density distribution.  
Another advantage of this method is that  
it is more straightforward to identify features in the density distribution that is not captured by the model which possibly indicate substructures in the Galaxy. 
{The result of this work also lay the foundation for future works to measure the global stellar population properties of the Milky Way, in a model-independent way, that allow direct comparison with observational results of external galaxies.}

In \textsection2, we illustrate the APOGEE data and age distribution of MAPs that are used to calculate the selection function. The calculation of the APOGEE selection function is then presented in the following \textsection3. In \textsection4 we present the intrinsic density distribution of MAPs after correcting for selection function, the density model, and the fitting results of structure parameters for each MAP. In \textsection5 we discuss the potential systematics in the results and compare our measurements with previous works. We also derive the density distribution and measure the structure of the total stellar populations in \textsection5.3 to facilitate comparison with earlier photometric works in the Milky Way and studies of external galaxies. Finally, a summary of our results is included in \textsection6. 

\section{Sample selection} 
\subsection{Selection criteria}
We select our stellar sample from APOGEE catalogue (allstar summary file) contained in the last internal data release after SDSS-IV Data Release 16 \citep[DR16;][]{ahumada2020,jonsson2020}, which includes data from observations until March 2020, reduced with a very slightly updated  version  of  the  DR16  pipeline  (r13).  APOGEE is a near-infrared, high-resolution spectroscopic survey \citep{blanton2017,majewski2017} that primarily targets evolved giant stars in the Milky Way and Local Group satellites \citep{zasowski2013,zasowski2017,beaton2021,santana2021}. This survey is performed using custom spectrographs \citep{wilson2019} with  the  2.5~m  Sloan  Telescope  and  the  NMSU 1~m Telescope at the Apache Point Observatory \citep{gunn2006,holtzman2010}, and with the 2.5~m Ir\'en\'ee du~Pont telescope at Las Campanas Observatory \citep{bowen1973}. 

The chemical abundances ([Fe/H], [Mg/Fe]) and stellar parameters (i.e., log$(g)$ and $T_{\rm eff}$) are taken from the APOGEE catalogue which are derived by custom pipelines ASPCAP described in \citet{nidever2015}, \citet{garcia2016} and Holtzman et al. in prep and line list in \citet{smith2021}. 

We use the recommended stellar ages and spectro-photometric distances in the astroNN Value Added Catalog for the internal data release \citep{mackereth2019}\footnote{The Value Added Catalog for the public Data Release 16 is available at https://data.sdss.org/sas/dr16/apogee/vac/apogee-astronn}. The ages are derived by training a deep neural network \citep{leung2019}\footnote{https://github.com/henrysky/astroNN} with asteroseismic ages derived from {\sl Kepler} and APOGEE combined observations \citep{pinsonneault2018}. We note that $\alpha$-rich stars at [Fe/H]$<-0.5$ may be subject to extra mixing \citep{shetrone2019} which would affect spectroscopic age determination but has not been considered in the astroNN age measurement. Moreover, the abundance space at [Fe/H]$<-0.5$, [Mg/Fe]$>0.2$ is not well populated by the training sample. Therefore, in this work we only use the astroNN ages for high-$\alpha$ ([Mg/Fe]$>0.2$) stars with [Fe/H]$>-0.5$. The distances in astroNN catalogue are determined by training the neural network on stars in common between {\sl Gaia} and APOGEE \citep{leung2019b}. 

We use the following criteria to select a sample of stars with reliable measurements from the parent APOGEE catalogue: 
\begin{itemize}
  \item Signal-to-Noise ratio (SNR) $>$ 50,
  \item EXTRATARG$==0$, 
  \item APOGEE\_TARGET1 and APOGEE2\_TARGET1 bit 9 $== 0$,
\end{itemize}
EXTRATARG==0 identifies stars in the main survey, in which stars were randomly selected, and removes duplicated observations. The ninth bit of APOGEE\_TARGET1 or APOGEE2\_TARGET1 are set for targets that are possible star cluster members. 
For more details about the APOGEE bitmasks, we refer the reader to \url{https://www.sdss.org/dr16/algorithms/bitmasks/}.

{To study the structural evolution of our Galaxy that consists of sub-populations formed at different epochs with different chemical abundances, it is ideal to obtain the density distribution of mono-age, mono-abundance abundances ([Fe/H] and [$\alpha$/Fe]) populations. However, as discussed in \textsection6.4, the number of observed stars in each APOGEE field is limited such that dissection on these three dimensions will result in too low statistics to recover robust intrinsic density. Since one motivation of this work is to provide density distribution of MAPs to allow direct measurement of global chemistry properties of the Milky Way, in this work we focus on the dimensions of [Fe/H] and [$\alpha$/Fe], which are well measured with comparable high precision.} We split the stellar sample into mono-abundance bins with bin size of 0.1~dex for [Mg/Fe] and 0.2~dex for [Fe/H]. Since these bin sizes are much larger than the typical abundance uncertainties in the catalogue (0.01--0.02~dex, \citep{poovelil2020}), there is more tolerance in the sample selection and we adopt a lower selection cut on SNR 
than other works using APOGEE data in order to include more stars. The final sample includes 301,634 stars in total. 

\subsection{Age distribution of mono-abundance populations}
\begin{figure*}
	\centering
	\includegraphics[width=18cm,viewport=0 40 1400 700,clip]{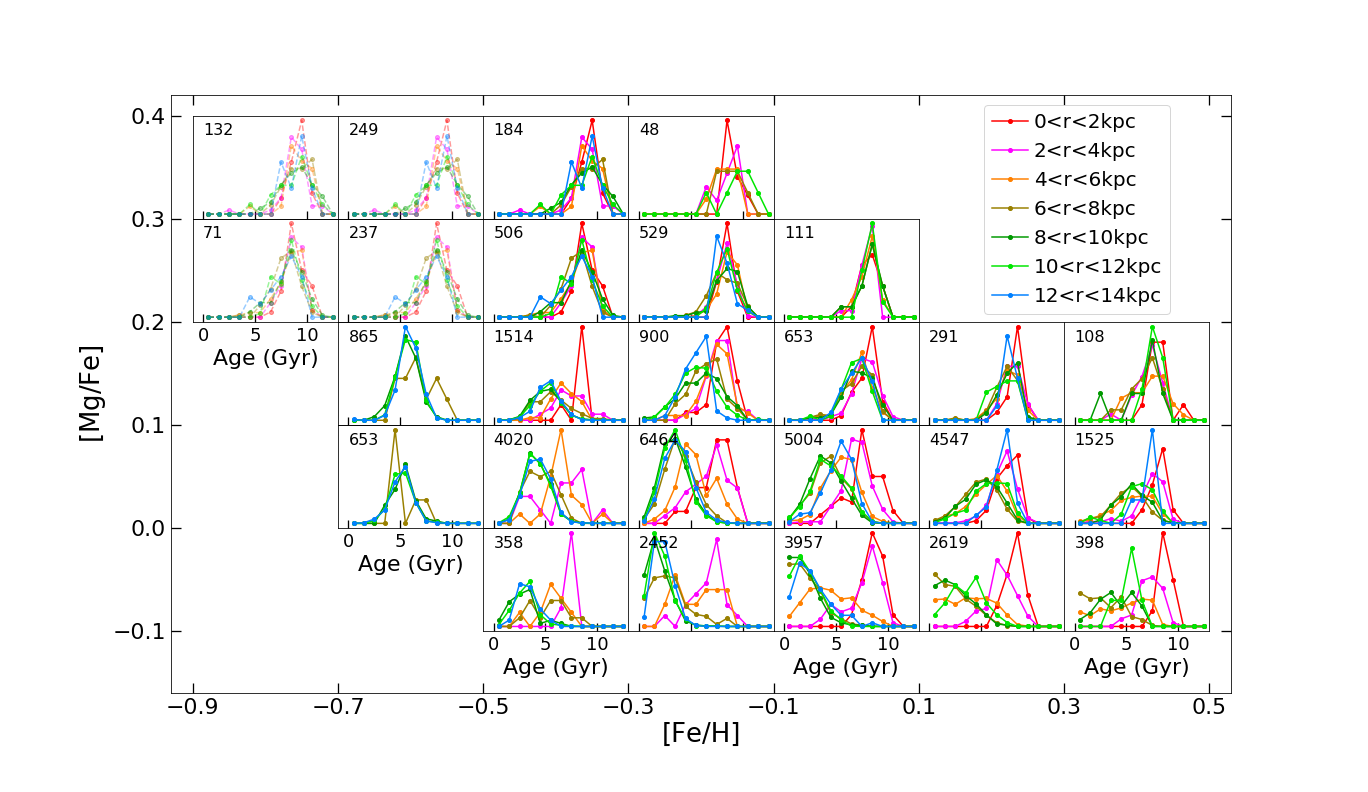}
	\caption{Normalized age distribution of mono-abundance populations on the mid-plane ($|{\rm Z}|<0.2$~kpc) as a function of Galactocentric radius. Since the age measurement at [Fe/H]$<-0.5$ is unreliable (see \textsection2.1), for the four high-$\alpha$ mono-abundance bins at [Fe/H]$<-0.5$ we assume they have the same age distribution of mono-abundance populations with the same [Mg/Fe] and $-0.5<$[Fe/H]$<-0.3$ (see \textsection2.2). The age distribution of these four MAPs are shown in dashed lines. The number at the left corner of each panel indicates the number of stars shown in each mono-abundance bin.        } 
	\label{age-dtr-r}
\end{figure*} 

\begin{figure*}
	\centering
	\includegraphics[width=18cm,viewport=0 40 1400 700,clip]{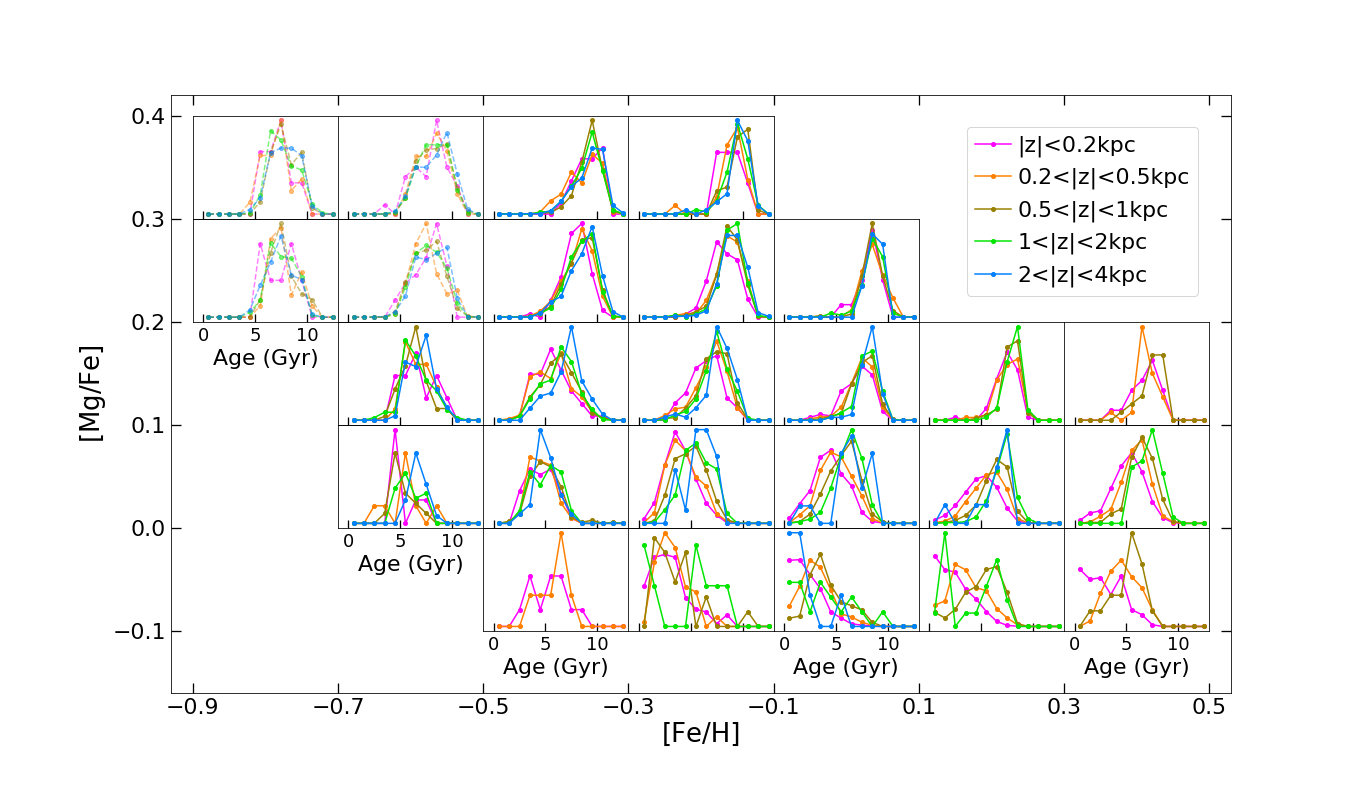}
	\caption{Similar to Fig.~\ref{age-dtr-r} but showing the age distribution as a function of $|{\rm Z}|$ height for mono-abundance populations at $6<{\rm R}<8$~kpc.  
	} 
	\label{age-dtr-z}
\end{figure*} 

To derive the effective selection function of APOGEE survey (i.e., fraction of the underlying population in the final spectroscopic sample) at a given [Fe/H] and [Mg/Fe], one prerequisite is the age distribution which essentially determines the distribution of the underlying population in the colour-magnitude diagram (CMD). 
Since the APOGEE targets are randomly selected from candidates well-defined in the CMD, we assume the age distribution of the observed sample in each MAP is representative of the underlying population. 
Figure~\ref{age-dtr-r} shows the normalized age distribution of MAPs as a function of Galactocentric radius for stars close to the disk plane ($|{\rm Z}|<0.2$~kpc). 

As we lack reliable age measurements at [Fe/H]$<-0.5$, to calculate the selection function for high-$\alpha$ ([Mg/Fe]$>$0.2) MAPs below this metallicity, we assume they have the same age distribution as the MAPs at the same [Mg/Fe] and $-0.5<$[Fe/H]$<-0.3$. This extrapolation is supported by the fact that the ages of stars in high-$\alpha$ sequence are similarly old \citep[e.g.,][]{feuillet2018}. Since the chemically defined thick disk (i.e., the high-$\alpha$ sequence) is generally believed to form rapidly at early times with a formation timescale around or less than 2~Gyr \citep[e.g.,][]{chiappini1997,kobayashi2006,lian2020b}, the age difference between stars in the high-$\alpha$ sequence should be within 2~Gyr. 
We have tested that, at age older than 8~Gyr, changing the age by 2~Gyr results in only a 7\% difference in the selected fraction of APOGEE targets. Therefore we conclude that the selection function of metal-poor, high-$\alpha$ mono-abundance bins derived from extrapolated age distributions is robust.  

It is interesting to note that the radial variation behavior in the age distribution is clearly different between MAPs (Figure~\ref{age-dtr-r}). In the high-$\alpha$ mono-abundance bins ([Mg/Fe]$>$0.2), the age distribution at various radii are rather similar, suggesting a spatially homogeneous chemically defined thick disk at the present day. This is in line with the nearly universal median locus of high-$\alpha$ sequence in the [$\alpha$/Fe]-[Fe/H] diagram \citep{nidever2014,hayden2015,weinberg2019,katz2021}. In contrast, the age distributions of MAPs at 
low-$\alpha$ ([Mg/Fe]$<$0.2) exhibit a clear and consistent trend with radius. Stars located at smaller radii tend to be older than their chemical counterparts at larger radii, i.e. negative age radial gradient in low-$\alpha$ MAPs. The strongest variation is seen in the solar abundance MAPs. These results suggest inhomogeneous chemical evolution and star formation histories during the formation of the thin disc. {The age distribution of MAPs with [Mg/Fe]=0.15, shown in the middle row of Fig. 1, present different trends with radius, from significant radial dependence at [Fe/H]$<$-0.1 to nearly no dependence at [Fe/H]$>$0.1, with a clear transition at solar [Fe/H]. Similarly, we can also see such transition at [Mg/Fe]=0.15 in the column of [Fe/H]=0. This transition reflects the separation of high- and low-$\alpha$ discs at [Mg/Fe] of $\sim0.15$ at solar [Fe/H] \citep{hayden2015}.}

Another implicit implication {of the radial variation of age distribution in low-$\alpha$ MAPs} is that the mixing effect from radial migration, if present, has not been strong enough to smooth out the difference in age and abundance distribution at different radii \citep{lian2022}. A disc formation scenario involving radially dependent gas accretion and dilution provides a natural explanation to the observed variation of age distribution in low-$\alpha$ MAPs \citep{haywood2019,lian2020a,lian2020b,lian2020c,katz2021}. In this scenario, the disc at larger radii has experienced more gas accretion and dilution, and therefore takes a longer time to enrich back to the same metallicity as the inner disc. Given the same starting time of gas accretion, the outer disc thus form younger stars at the same metallicity than the inner disc. 

Similar to Fig.~\ref{age-dtr-r}, Figure~\ref{age-dtr-z} shows the normalized age distribution of MAPs as a function of vertical height above the disc plane for stars at $6<{\rm R}<8$~kpc. Similar to the trend with radius, the age distributions of MAPs at [Mg/Fe]$>0.2$ show little variation with height, while the low-$\alpha$ stars with the same abundances tend to be older at larger vertical distance. A possible origin of this vertical trend is secular disc heating, possibly by the pre-existing thick disc or interactions with satellite galaxies \citep[e.g.,][]{mackereth2019}. Note that the age 
variation of MAPs within the vertical extent of the disc is clearly smaller than within the radial extent. This implies that the radially varying star formation history may have played a more important role in shaping the Milky Way's thin disc than other evolutionary processes.   

\section{APOGEE selection function}

We derive the effective selection function of APOGEE survey in three steps using the last internal release after DR16. The first one is to estimate the fraction of stars that fall within the candidate selection criteria of APOGEE,  which is explained in \textsection3.1. Since the target candidates are selected in near-infrared CMD, we hereafter refer to this fraction as $f_{\rm CMD}$. The next step is to calculate the fraction of target candidates that are indeed observed in the main survey, $f_{\rm spec}$. Finally, the last step is to measure the fraction of targets that is included in the final stellar sample that have reliable measurements of abundances (as discussed in \textsection2.1), $f_{\rm abun}$. The calculation of the latter two are described in \textsection3.2. 

\subsection{Target candidates selection}
\begin{figure}
	\centering
	\includegraphics[width=8cm]{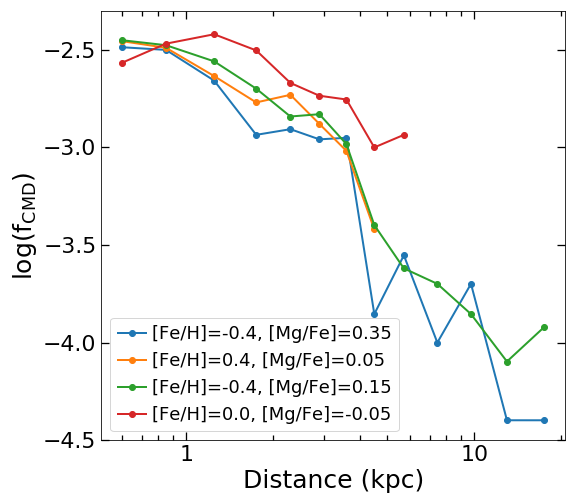}
	\caption{Fraction of the underlying stellar population satisfying APOGEE target candidates definition 
	as a function of heliocentric distance for a short cohort in a APOGEE disc field (field ID: 2544, $l=-24.6\degree$, and $b=46.5\degree$). The fraction of mono-abundance populations are calculated separately and four of them are shown in four lines with different colour. 
	} 
	\label{fcmd}
\end{figure} 

The targets in the APOGEE main survey are selected from candidates defined by $(J-K_{\rm})_0$ colour and $H$ magnitude in 2MASS Point Source Catalog \citep{skrutskie2006}. For full details of the targeting strategy, we refer the reader to \citet{beaton2021}
and \citet{santana2021}, 
but we summarize the relevant information here. The dereddened $(J-K_{\rm})_0$ colour is derived using the Rayleigh-Jeans Color Excess Method \citep{majewski2011} or extinction map from \citet{schlegel1998}. The range of $(J-K_{\rm})_0$ colour for target selection varies between fields. APOGEE-1 and APOGEE-2 bulge fields use a single colour limit of $(J-K_{\rm})_0\ge0.5$, while APOGEE-2 halo fields adopt a bluer cut with $(J-K_{\rm})_0\ge0.3$. To increase the fraction of distant red giant stars, APOGEE-2 disc fields select targets separately from two colour bins; $0.5\le(J-K_{\rm})_0<0.8$ and $(J-K_{\rm})_0\ge0.8$. 
In each field, a group of stars with exactly the same visits are referred to as a `cohort'. For a single field, there are a maximum of three cohorts, with three non-overlapping $H$ magnitude ranges corresponding to three different depth of observations. While the exact range of $H$ magnitude used for target selection depends on the field and number of visits of the cohort, for reference, the typical $H$ magnitude range for the most common `short' cohort which has the least number of visits is 7$-$12.2~mag. 

For each MAP at each heliocentric distance in a given field and cohort, we use the galactic location-dependent age distribution and PARSEC evolution tracks\footnote{http://stev.oapd.inaf.it/cgi-bin/cmd} \citep{bressan2012} with bolometric corrections from \citet{chen2019} to simulate the stellar distribution in the observed $(J-K_{\rm})_0$-$H$ CMD. 
We adopt a Kroupa IMF \citep{kroupa2001} and calculate extinction in $H$ band using the python package mwdust\footnote{https://github.com/jobovy/mwdust} \citep{bovy2015a} and a combined 3D dust map from \citet{marshall2006}, \citet{green2019}, and \citet{drimmel2003} (see \citealt{bovy2015a} for details on the combination). 
The first component of APOGEE selection function, $f_{\rm CMD}$, 
is then calculated by applying the APOGEE target selection limits and counting the fraction of stars that falls within these limits. 


Figure~\ref{fcmd} illustrates this fraction as a function of heliocentric distance for a short cohort in a disk field (field ID: 2254) as an example. Each line indicates a single mono-abundance bin as denoted in the legend. 
Note that the MAPs in the legend from top to bottom have decreasing mean ages. It can be seen that $f_{\rm CMD}$ generally decreases with increasing distance, mainly due to fainter apparent $H$ magnitude. Three of the four mono-abundance bins have a similar $f_{\rm CMD}$, while the youngest one ([Fe/H]=0, [Mg/Fe]=-0.05) shows lower fraction at the smallest distance but higher fraction at distance beyond 3~kpc. This different behavior at various distances is a result of two competing effects: younger populations are brighter, which increases their $f_{\rm CMD}$, but at younger ages the red giant stage is populated by more massive stars, which are rarer and spend less time in the RGB, resulting in a lower $f_{\rm CMD}$. At small distances, red giant stars are generally bright enough to meet the target candidate selection limits and therefore the latter effect is dominant. In contrast, at larger distances, the brightness of stars become more important and younger stars will have larger chance to be selected as targets. However this age effect is only significant at young ages when the age-luminosity relation is relatively steep. Therefore the selected fraction of the three mono-abundance bins with intermediate and old ages do not show clear deviation from each other.

\subsection{Selection of spectroscopic target and final abundance sample}
\begin{figure}
	\centering
	\includegraphics[width=8.7cm]{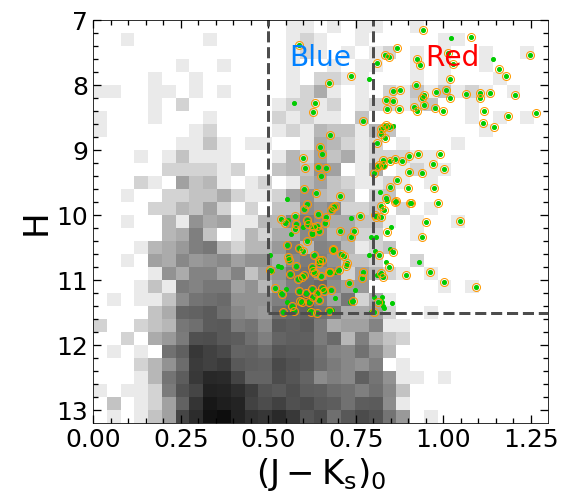}
	\caption{APOGEE target selection in ${\rm (J-K_{\rm s})_0-H}$ for the same field and cohort as Fig.~\ref{fcmd}. The distribution of stars in 2MASS Point Source Catalogue in this field is shown in the background in grey. 
	Green dots are those that are selected for targeting and the orange circles denote stars in the final stellar sample with robust abundance measurements. Dashed lines indicate the selection limits for target candidates in the CMD.  
	For this field the target selection is performed in two colour bins separately to have a better coverage of distant stars.  
	} 
	\label{select-spec}
\end{figure} 

\begin{figure}
	\centering
	\includegraphics[width=8.5cm]{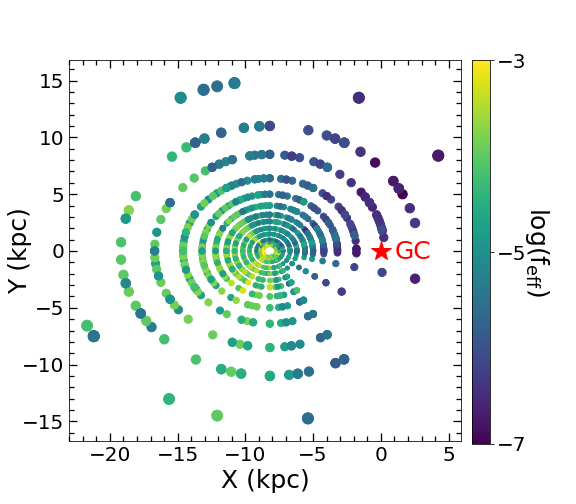}
	\caption{Distribution of effective selection fraction at various distances of all cohorts with $|{\rm Z}|<0.2$~kpc in X-Y plane. 
	} 
	\label{select-final}
\end{figure} 

APOGEE main survey targets are randomly drawn from the photometric candidates defined in $(J-K_{\rm})_0$ and $H$. 
The selected fraction in this step, $f_{\rm spec}$, is a only function of field and cohort, and also $(J-K_{\rm})_0$ colours in case of APOGEE-2 disc fields. The $f_{\rm spec}$ for APOGEE DR16 is publicly available\footnote{https://github.com/jobovy/apogee} \citep{bovy2014,bovy2016_open}. We assume this fraction does not change from DR16 to the incremental post-DR16 release and adopt this $f_{\rm spec}$ in this work. 

In some unusual cases, such as abnormally low spectra quality/SNR, extreme stellar parameters close to or exceeding the edge of ASPCAP grid, a small fraction of targeted stars do not have well-determined chemical compositions. These objects are excluded from the final stellar sample by applying the selection criteria described in \textsection2.1. 
For each MAP in each field and cohort, we calculate the fraction of targeted stars meeting the selection criteria in \textsection2.1 for reliable abundance measurements, $f_{\rm abun}$, which is the final patch of the APOGEE effective selection function.  
Figure~\ref{select-spec} illustrate the selection of targets and final stellar sample for the same field and cohort in Fig.~\ref{fcmd}. 
The distribution of stars in the 2MASS Point Source Catalogue in this field is shown in grey in the background. Selected targets are indicated as green dots, with $f_{\rm spec}$ in the blue and red colour bins of 0.166 and 0.853, respectively. Stars in the final stellar sample are shown in orange circles, which comprise 97.3\% and 92.0\% of the targeted sample in the blue and red colour bins, respectively. 
The effective selection fraction, $f_{\rm eff}$, is a product of the fraction in the three selection steps as:
\begin{equation}
\begin{split}
f_{\rm eff}({\rm [Fe/H],[Mg/Fe],F,C,}d) & = \\
 & f_{\rm CMD}({\rm [Fe/H],[Mg/Fe],F,C},d)\\
 & \times f_{\rm spec}({\rm F,C})\times f_{\rm abun}({\rm F,C}) \\
\end{split} 
\end{equation}
for fields have target candidates selected on single $J-K_{\rm 0}$ colour bin, or 
\begin{equation}
\begin{split}
f_{\rm eff}({\rm [Fe/H],[Mg/Fe],F,C}&,J-K_{\rm 0},d)  =\\
&f_{\rm CMD}({\rm [Fe/H],[Mg/Fe],F,C},J-K_{\rm 0},d)\\
&\times f_{\rm spec}({\rm F,C},J-K_{\rm 0})\\
&\times f_{\rm abun}({\rm F,C},J-K_{\rm 0})\\
\end{split} 
\end{equation}
for fields with target candidates selected in two colour bins separately. Here $d$ stands for heliocentric distance, F and C stand for field and cohort, respectively. To calculate a consistent $f_{\rm eff}$ for all fields, in fields with targets selected in two colour bins, we further calculate an average $f_{\rm eff}$ weighted by the number of stars in each colour bin as predicted by the theoretical isochrones in \textsection3.1. 
Note that, to calculate the intrinsic number density, 
dividing the total number of stars observed in the two colour bins by the average selection fraction is equivalent to doing this calculation separately in each colour bin and then summing them up. 
For reference, $f_{\rm eff}$ for the low-$\alpha$, metal-rich MAP of [Mg/Fe]$=$0.05 and [Fe/H]$=$0.4 at a distance of 1~kpc in the short cohort of the field in Fig.\ref{select-spec} are 6.30$\times10^{-4}$ and 3.06$\times10^{-3}$ in the blue and red colour bins, respectively.  

To give a global view of the APOGEE effective selection function, we show in Figure~\ref{select-final} the X-Y distribution of $f_{\rm eff}$ of all cohorts in the Galactic plane ($|{\rm Z}|<0.2$~kpc). 
Each cohort is separated into 13 distance bins, from 0.5 to 20~kpc. The bin size is smaller for closer distances to ensure approximately even number of stars in each distance bin. A more detailed discussion on the number statistics is given in $\textsection$6.1.   
The effective selection fraction is higher at smaller distance where stars are brighter, and lower in the Galactic center direction owing to higher stellar density and extinction towards the inner Galaxy.

\section{Density distribution of underlying mono-abundance populations}
\begin{figure*}
	\centering
	\includegraphics[width=18cm,viewport=10 100 2300 1150,clip]{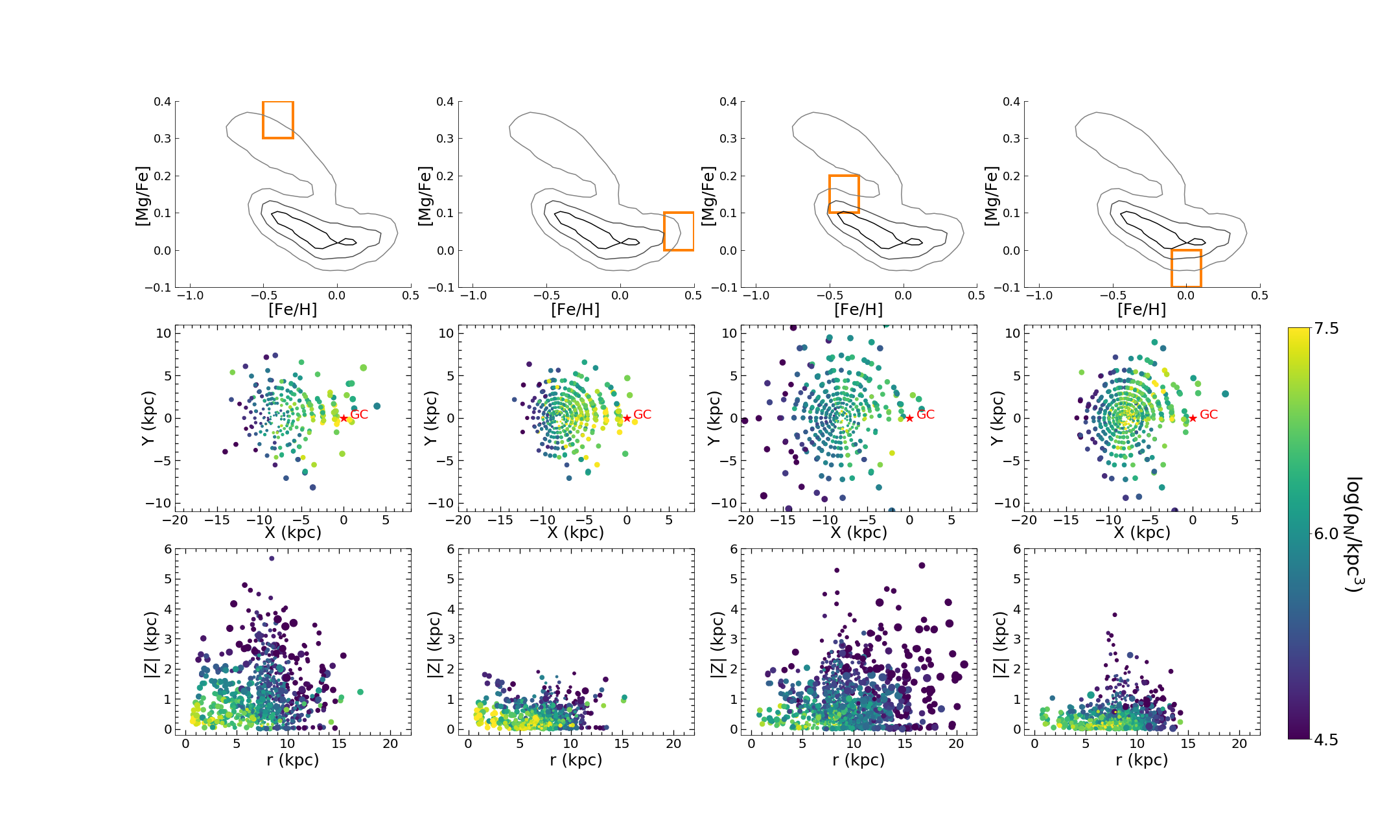}
	\caption{Spatial distribution of intrinsic number density of four MAPs in the X-Y plane (with $|{\rm Z}|<0.2$~kpc) in the middle row and the R-$|{\rm Z}|$ plane in the bottom row. Each symbol indicates one spatial bin, with the location determined by the average position of observed stars in each bin. The symbol size of each bin is proportional to its heliocentric distance. Each column is for one MAP, with decreasing mean age from the left to the right. The abundance range of each MAP is indicated by the orange box in the top row, on top of the distribution of the whole APOGEE sample shown in grey contours. 
	} 
	\label{xy-rz}
\end{figure*} 

\subsection{Spatial distribution of sampled underlying populations}
With the effective selection function, we can derive the number density of underlying populations by simply dividing the observed density by the selected fraction:
\begin{equation}
    \rho_{\rm int}({\rm [Fe/H],[Mg/Fe], F, C},d) = \frac{\rho_{\rm obs}({\rm [Fe/H],[Mg/Fe], F,C},d)}{f_{\rm eff}({\rm [Fe/H],[Mg/Fe], F, C},d)},\\
\end{equation}
where $\rho_{\rm obs}$ indicates the observed number density, which equals the number of observed stars divided by the volume of each distance bin at a given field. The different fields of view on the telescopes at Apache Point Observatory and Las Campanas Observatory have been taken into account. 
The calculation of intrinsic number density is performed independently for each combination of [Fe/H], [Mg/Fe], and distance in each field's set of cohorts. 
To improve statistics, we rebin all APOGEE fields with a semi-regular ($l, b$) grid with 2.7 fields in each grid element on average. 
For each spatial location ($l, b, d$), the average recovered intrinsic number density of all cohorts is used.  

Figure~\ref{xy-rz} shows the spatial distribution of intrinsic number density distribution of four MAPs in the X-Y plane (with $|{\rm Z}|<0.2$~kpc) in the middle row and the R-$|{\rm Z}|$ plane in the bottom row. 
Each symbol indicates one spatial bin, with the location determined by the average position of observed stars within this bin. The symbol sizes are proportional to their distance to the Sun. Each column shows the distribution of a single MAP formed at different epochs of the Milky Way's evolutionary history, following an order of decreasing mean age from the left to the right. In the top row, the position of each MAP in [Mg/Fe]-[Fe/H] distribution is shown as orange box, on top of the whole APOGEE sample shown as grey contours.

It is interesting to note that stellar populations with different abundances have {noticeably different three dimensional spatial distributions}. 
The old, high-$\alpha$ populations {(first column in Fig.~\ref{xy-rz}) have relatively limited radial distribution but rather extended vertical distribution.} 
This is consistent with the previously-established thick and compact morphology of the chemically defined thick disc \citep{bovy2012b,bovy2016}. 
The metal-rich, low-$\alpha$ stars (second column in Fig.~\ref{xy-rz}) are believed to form 
following the high-$\alpha$ population, possibly after a rapid early star formation quenching episode \citep{haywood2018,lian2020b,lian2020c,khoperskov2021}. These stars have a similar spatial extent in radius but a shorter vertical extent (i.e., thinner) compared to the high-$\alpha$ population. The similarity in radial extension supports the hypothesis of an evolutionary connection between these two stellar popoulations. The discrepancy in vertical extension, {combined with their distinct chemical abundances \citep[e.g.,][]{hayden2015} and different kinematic properties \citep[e.g.,][]{robin2017},} however, suggests an upside-down disc formation with a transition from rapid assembly of a kinematically hot, thick disc to secular establishment of a kinematically cold, thin disc \citep{bird2013,freudenburg2017}. This transition is likely connected to the early quenching process, possibly driven by the same mechanism, which remains unclear. 

The metal-poor, low-$\alpha$ population (third column in Fig.~\ref{xy-rz}), which has an average age slightly younger than the metal-rich, low-$\alpha$ population, displays a very different spatial distribution. It shows a clearly extended distribution in both radial and vertical direction. In particular, among the four MAPs considered, this metal-poor, low-$\alpha$ MAP shows the largest radial distribution, out to 20~kpc. This is in line with the finding that the outer disc beyond 15~kpc is dominated by metal-poor, low-$\alpha$ stars (\citealt{lian2022}, Evans et al. in prep). The large vertical extension of this population, combined with its wide radial distribution, 
leave the metal-poor, low-$\alpha$ stars as the dominant population in the geometrical thick disc at large radii. The similar vertical extension between the high-$\alpha$ and metal-poor, low-$\alpha$ populations suggest they were both born in a hot kinematic environment but at different times. This is consistent with a star burst origin of the metal-poor, low-$\alpha$ stars triggered by gas accretion and possible interaction with infalling satellite \citep{buck2020,lian2020a,lian2020b,agertz2021}. 
The transition of the dominant population at large $|{\rm Z}|$ height from inner to outer Galaxy gives rise to the reported radial gradient of age and abundances in the geometric thick disc \citep{boeche2013,boeche2014,martig2016,lian2020b}.         

The youngest population in our sample has solar-like abundances, shown in the fourth column in Fig.~\ref{xy-rz}. It can be seen that they have moderate radial extension, wider than the two MAPs in the left but less extended than the metal-poor, low-$\alpha$ MAP, and the shortest vertical extension. This suggests that the recent star formation in the Milky Way over the past 1--2~Gyr has occurred mostly at intermediate radius and strictly confined to the disc plane. 

\subsection{Radial and vertical density profiles of mono-abundance populations}
\begin{figure*}
	\centering
	\includegraphics[width=18cm,viewport=0 40 1400 700,clip]{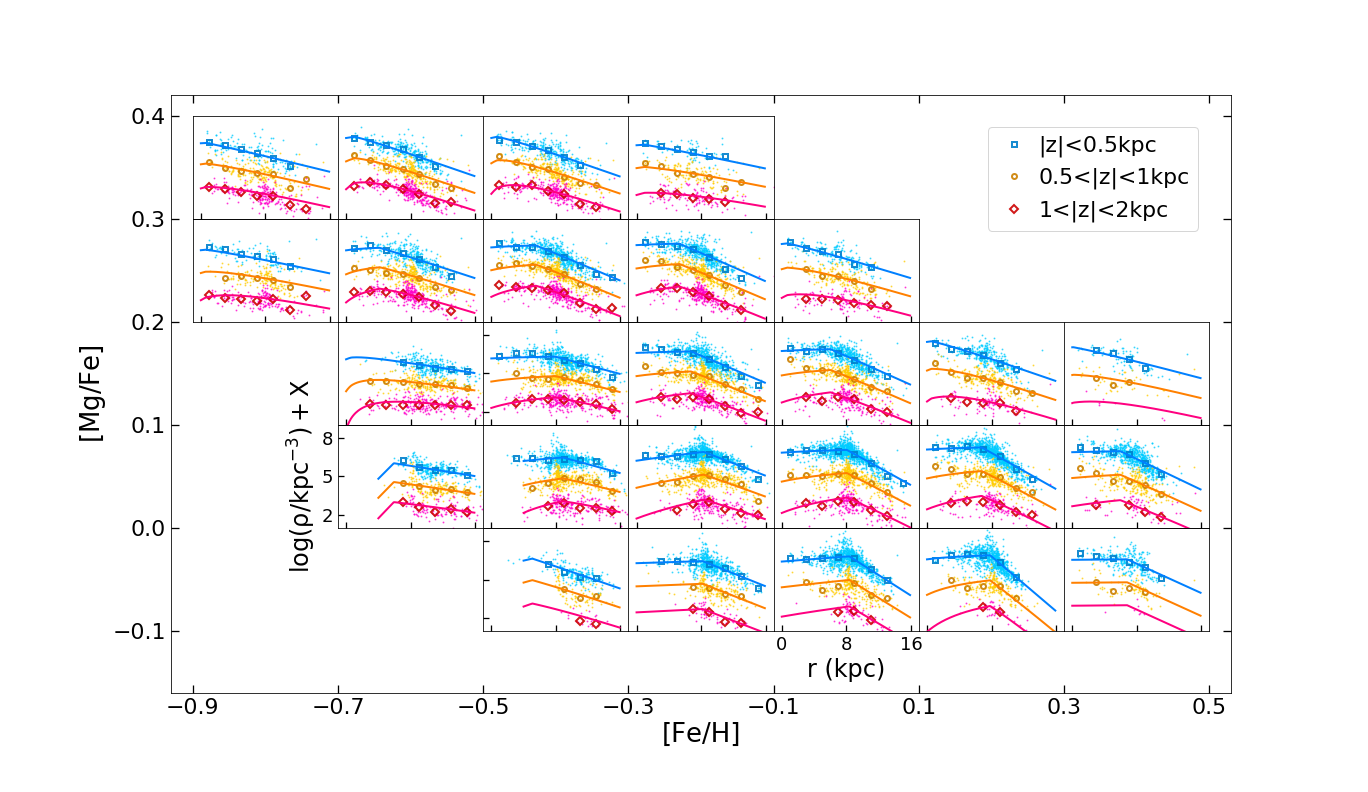}
	\caption{Radial density profile of mono-abundance populations at three $|{\rm Z}|$ height bins as illustrated in the top-right legend. Small dots indicate intrinsic density measurements at each location. Enlarged symbols show the median intrinsic density at various radial bins with bin size of 2~kpc. To better visualize the profile at different heights, an arbitrary negative shift in y-axis is added to the two height bins above the disc plane. The solid lines represent the best-fitted density model that will be discussed in \textsection5.1 below. 
	} 
	\label{den-rprof}
\end{figure*} 

\begin{figure*}
	\centering
	\includegraphics[width=18cm,viewport=0 40 1400 700,clip]{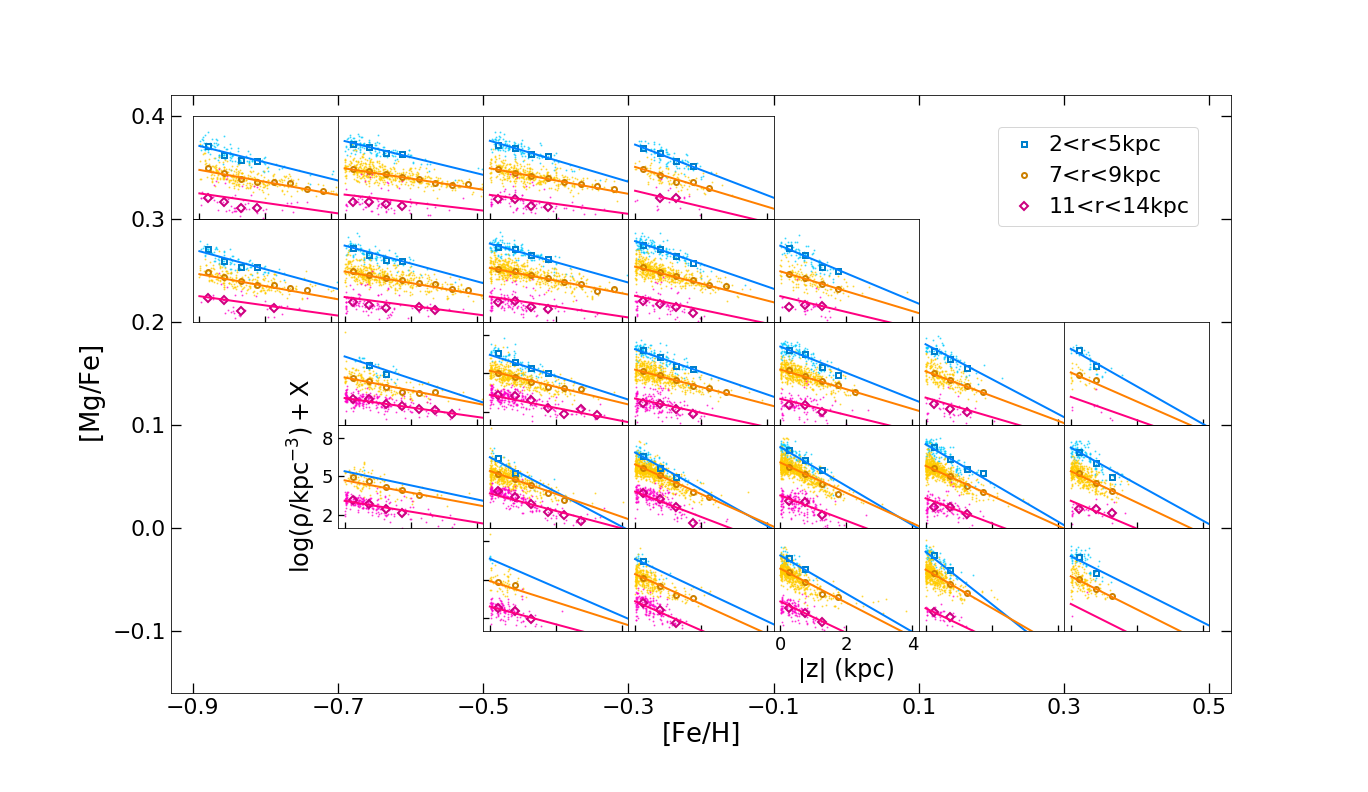}
	\caption{Similar to Fig.~\ref{den-rprof} but showing the vertical density profile at three radial bins. An arbitrary negative shift in y-axis is added to the two radial bins in orange and magenta. 
	} 
	\label{den-zprof}
\end{figure*} 

Here we perform a comprehensive analysis of the intrinsic, selection-function corrected, number density distribution of underlying population as a function of Galaxy location (R, $|{\rm Z}|$) and chemical abundances ([Fe/H], [Mg/Fe]).  

Figure~\ref{den-rprof} (Figure~\ref{den-zprof}) shows the radial (vertical) density profile of MAPs at three different heights (radii). For visual clarity, an arbitrary density offset is added to the data points denoted in orange and magenta dots. Enlarged squares represent the median number density as a function of radius or height, and are included to highlight the shape of density profiles. 

Surprisingly, most MAPs, including many high-$\alpha$ MAPs with [Mg/Fe]$>0.2$, display a non-linear shape in their radial density distribution in logarithm, with a break near the solar radius and a flatter distribution in the inner Galaxy. 
This non-linear profile shape seems to be present in the radial density profile of stars with different colours in \citet{juric2008}, which was interpreted as a wide overdensity feature. A broken radial density distribution with two exponential profiles was used by \citet{bovy2016} and \citet{mackereth2017} to fit the APOGEE data {and by \citet{yu2021} for LAMOST data}. With the latest APOGEE observations, 
we directly recover the density distribution of underlying populations and confirm the presence of a broken radial density profile. 
Given the radial extent of the `flat' part of the distributions, 
it seems that the broken radial profile is unlikely to be caused by an overdensity structure in the disc. {It is always possible that the profile density break near the solar radius measured here and by others is due to some observational bias. However, there are also reasons that a break here is physically plausible. For example, a} possible origin of this broken radial profile is the presence of outer Lindbald Resonance (OLR), which limits churning migration {across this radius and therefore} separate the disc into two parts with little exchange as found in the simulation by \citet{halle2015}. It is suggested that the current position of the OLR of the Milky Way is slightly inside the solar radius \citep{dehnen2000,famaey2005,minchev2007}, close to the position of the break seen in the radial density profile. It is also interesting to note that the chemical properties of the outer disc ($R>10$~kpc) are substantially different from the inner disc \citep[e.g.,][]{haywood2013,anders2014,nidever2014}, which may also be caused by the existence of OLR \citep{halle2015}. 

{It is interesting to note that} in external disc galaxies with deep imaging data, a broken radial density profile with two distinct exponential components is frequently seen \citep[e.g.,][]{pohlen2002,pohlen2006}. For example, \citet{pohlen2002} found galaxies that exhibit a broken radial profile with a shallow inner and steeper outer exponential region, qualitatively consistent with the pattern we report here in the Milky Way. \citet{pohlen2006} studied a sample of 90 nearby spiral galaxies and reported a fraction of 60\% of these galaxies showing such broken profile which was referred to as `downbending' break. Another 30\% galaxies exhibiting a different broken profile (i.e., steep inner and shallower outer region) and only 10\% galaxies have a pure exponential disc down to the surface brightness limit. Using a set of hydrodynamic simulations of disc galaxy formation, \citet{herpich2015} found an intriguing connection between the type of break in galaxy profile and the host halo's initial angular momentum. In particular, galaxies with high angular momentum in their halo tend to have downbending broken profiles.       

The vertical density distribution of mono-abundance populations shown in Fig.~\ref{den-zprof} can be generally described by a single exponential profile. The distribution seems to flatten at large vertical distance ($|{\rm Z}|>2$~kpc) in some MAPs (e.g., [Fe/H]$=$-0.4, [Mg/Fe]=0.05), but more observations far off the disc will be needed to draw a more affirmtive conclusion. The vertical density distribution tend to be flatter with increasing radius in many MAPs, which is usually referred to as the `flaring' feature in the disc (\textsection4.3). 

\subsection{Radial variation of scale height}
\begin{figure*}
	\centering
	\includegraphics[width=18cm,viewport=0 40 1400 700,clip]{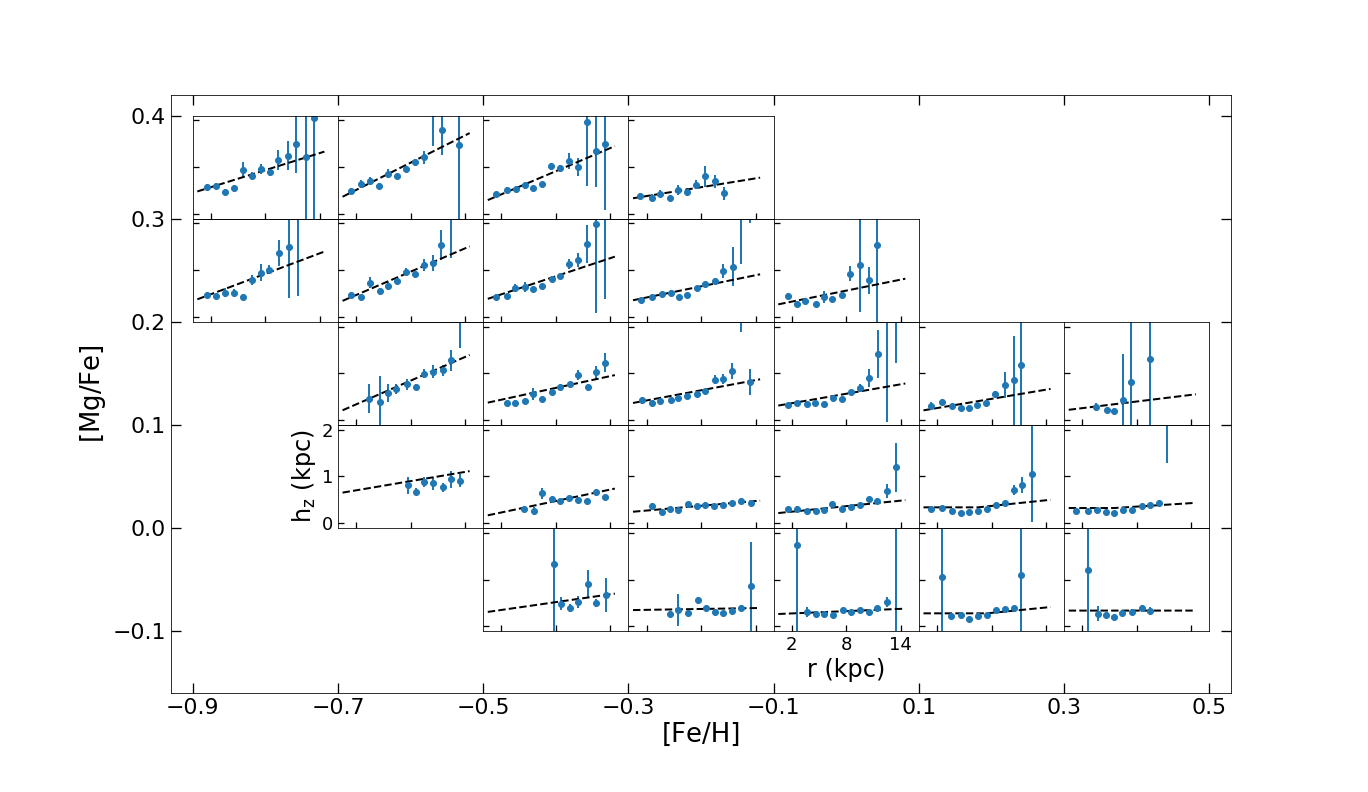}
	\caption{Scale height of mono-abundance populations as a function of radius. Error bars indicate the uncertainties of scale height estimated via Monte Carlo simulation. Dashed lines indicate the prediction of 2D density model (\textsection5.1).   
	} 
	\label{hz-r-simple}
\end{figure*} 

Before performing a full parametric fitting to the density distribution $\rho({\rm R,|Z|})$, we first measure the scale height in a series of narrow radial bins from 0 to 15~kpc ($\Delta R=1$~kpc). A single exponential profile is used to fit the vertical density distribution up to 4~kpc in each radial bin. Figure~\ref{hz-r-simple} shows the best-fitted scale height as a function of radius for each 
MAP. Most MAPs, except for those with [Mg/Fe]$<0$, 
exhibit a clear trend of increasing scale height (i.e. thickness) with radius (i.e., flare). 
Surprisingly, the high-$\alpha$ populations display the steepest increase of scale height with radius, i.e., the strongest flaring. This is at odds with the finding in \citet{bovy2016} of a constant thickness of high-$\alpha$ stars, but more consistent with the recent work by \citet{yu2021} using LAMOST data. \citet{yu2021} showed that both the high- and low-$\alpha$ MAPs flare with comparable strength. Another interesting finding in \citet{yu2021} is that the flaring in low-$\alpha$ MAPs occurs mostly at ${\rm R}>10$~kpc.  
Such flaring pattern of low-$\alpha$ stars seems also present in some low-$\alpha$ MAPs in Fig.~\ref{hz-r-simple} (e.g., MAP at [Fe/H]=-0.4 and [Mg/Fe]=0.15). 

\section{Structure of mono-abundance populations}
To quantitatively compare the structure between various MAPs, we perform a 2D parametric fitting to the intrinsic density distribution of each MAP, in radius and vertical distance simultaneously. 

\subsection{Density model}
The density model adopted from \citet{bovy2016} has also been used to derive the structure parameters of mono-age or mono-abundance populations in many other works \citep[e.g.,][]{mackereth2017,yu2021}.  
It consists of two components, which describes the density distribution in radial and vertical direction separately.
\begin{equation}
    \rho(R,|Z|) \equiv \upsilon(R)\zeta(R,|Z|),\\
\end{equation}

The radial component is a broken exponential profile with three free parameters: break radius (${R_{\rm b}}$), and the scale length inside and outside the break radius ($h_{\rm R,inner}$ and $h_{\rm R,outer}$).    

    \[ \upsilon(R) \propto  \begin{cases} 
      e^{R/h_{\rm R,inner}} & R\leq R_{\rm b} \\
      e^{-R/h_{\rm R,outerr}} & R>R_{\rm b} \\
   \end{cases}
     \]   
     

The vertical component in the density model is a single exponential profile. 
Based on the analysis in \textsection4.3 and Fig.~\ref{hz-r-simple}, we set the scale height as a linear function of radius as following:
\begin{equation}
    \zeta(R, |Z|) \propto e^{-|Z|/h_{\rm Z}(R)},\\
    h_{\rm Z}(R) = A_{\rm flare}\times (R-R_{\odot})+h_{\rm Z,R_{\odot}}
\end{equation}
The slope of this linear function ($A_{\rm flare}$) represents the strength of flaring, and the intercept is the scale height at the solar radius, which are both set to be free. 
The density model is scaled to the number density at the solar radius in the disc plane, which is the last free parameter in the model. In this work, we adopt a solar position in the Galaxy of $R_{\odot}=8.2$~kpc and $Z=0.027$~kpc \citep{bland2016}.  

To summarize, there are six free parameters in the density model adopted in this work:
\begin{itemize}
  \item $\rho_{\odot}$, intrinsic density at solar radius on the disc plane, 
  \item $R_{\rm b}$, break radius in the radial density distribution,
  \item $h_{\rm R,inner}$, scale length at $R\leq R_{\rm b}$,
  \item $h_{\rm R,outer}$, scale length at $R>R_{\rm b}$,
  \item $A_{\rm flare}$, slope of radial variation in scale height,
  \item $h_{\rm Z,R_{\odot}}$, scale height at solar radius.
\end{itemize}
We fit the density model simultaneously to all the spatial intrinsic density measurements in a given MAP using \texttt{curve\_fit} function in the optimize module of SciPy \citep{virtanen2020}. 




The solid lines in Fig.~\ref{den-rprof} and Fig.~\ref{den-zprof} show the the predicted 1D density profile of best-fitted models, which is sliced from the 2D model at the median height or radius of the density measurements from which the observed median profiles are drawn. For example, to compare with the observed median radial density profile at $|{\rm Z}|<$0.5~kpc, the predicted radial density profile is a slice of the 2D model at the median $|{\rm Z}|$ of density measurements within $|{\rm Z}|<$0.5~kpc.  

The best-fitted models well match the intrinsic density distribution of MAPs in both radial and vertical direction. 
For example, the broken radial density profile of low-$\alpha$ MAPs are well reproduced by the model with double exponential components in radius. 
Note that individual intrinsic density measurements are more frequent close to Sun because of the adopted binning in distance. The good match between the model prediction and the observed median radial and vertical density profiles suggests that the concentration of density measurements at solar vicinity does not significantly affect the result. 

There are deviations in some MAPs, which might suggest interesting sub-structures that are not included in the smooth density model. For instance, in many low-$\alpha$ MAPs, the intrinsic number density is higher than predicted close to Galactic center (${\rm R<4}$~kpc). 
This feature is not captured by the model and may be signature of the bar in the inner Galaxy. 

The predicted radial variation of scale height of the 2D density model is shown as dashed line in Fig~\ref{hz-r-simple}. It can be seen that it matches well the scale height derived from 1D fitting in narrow radial bins, confirming the robustness of the 2D density fitting approach used in this work. The interesting non-linear radial variation of scale height in some low-$\alpha$ MAPs is not reproduced by our 2D model which assumes linear flaring. To reproduce this feature requires a different (e.g., exponential flaring in \citet{bovy2016} and \citet{mackereth2017}) or a more complex (e.g., step-wise) flaring disc model which is beyond the scope of this work. Nonetheless, this non-linear feature may provide valuable insights into the origin of disc flaring and worth being studied in more detail in the future.   

\subsection{Structure parameters of number density distribution}
\begin{table*}
    \caption{Best-fitted structure parameters of individual MAPs.}
    \centering
    \begin{tabular}{l c c c c c  }
    \hline\hline 
    [Fe/H], [Mg/Fe] & $\rho_{\rm \odot, N}$ & $R_{\rm b}$ & $h_{\rm R,outer}$ & $A_{\rm flare}$ & $h_{\rm Z,R=R_{\odot}}$  \\
     & log($\rm Number/kpc^{3}$) & kpc & kpc & - & kpc  \\
    \hline
-0.4,-0.05 & 6.149$\pm$0.066 & 5.09$\pm$0.816 & 1.961$\pm$0.178 & 0.028$\pm$0.02 & 0.525$\pm$0.074 \\
-0.2,-0.05 & 6.662$\pm$0.03 & 8.193$\pm$0.107 & 1.722$\pm$0.075 & 0.003$\pm$0.005 & 0.374$\pm$0.014 \\
0.0,-0.05 & 7.049$\pm$0.025 & 8.366$\pm$0.074 & 1.075$\pm$0.035 & 0.008$\pm$0.003 & 0.328$\pm$0.009 \\
0.2,-0.05 & 6.961$\pm$0.07 & 7.826$\pm$0.25 & 0.8$\pm$0.044 & 0.019$\pm$0.017 & 0.293$\pm$0.084 \\
0.4,-0.05 & 6.432$\pm$0.067 & 6.788$\pm$1.124 & 1.525$\pm$0.268 & 0.0$\pm$0.007 & 0.339$\pm$0.03 \\
-0.6,0.05 & 5.911$\pm$0.033 & 5.921$\pm$0.496 & 4.232$\pm$0.282 & 0.033$\pm$0.009 & 0.908$\pm$0.066 \\
-0.4,0.05 & 6.631$\pm$0.089 & 8.85$\pm$1.532 & 2.42$\pm$0.498 & 0.041$\pm$0.007 & 0.485$\pm$0.019 \\
-0.2,0.05 & 7.142$\pm$0.024 & 8.317$\pm$0.096 & 1.743$\pm$0.05 & 0.017$\pm$0.002 & 0.378$\pm$0.007 \\
0.0,0.05 & 7.253$\pm$0.015 & 8.06$\pm$0.042 & 1.23$\pm$0.025 & 0.02$\pm$0.001 & 0.371$\pm$0.007 \\
0.2,0.05 & 6.993$\pm$0.018 & 6.776$\pm$0.079 & 1.206$\pm$0.03 & 0.02$\pm$0.002 & 0.376$\pm$0.01 \\
0.4,0.05 & 6.554$\pm$0.034 & 5.956$\pm$2.408 & 1.42$\pm$0.165 & 0.012$\pm$0.007 & 0.36$\pm$0.013 \\
-0.6,0.15 & 5.886$\pm$0.028 & 0.643$\pm$2.158 & 4.225$\pm$0.194 & 0.084$\pm$0.006 & 0.862$\pm$0.035 \\
-0.4,0.15 & 6.396$\pm$0.032 & 7.935$\pm$0.77 & 2.511$\pm$0.12 & 0.042$\pm$0.004 & 0.699$\pm$0.018 \\
-0.2,0.15 & 6.477$\pm$0.025 & 6.737$\pm$0.231 & 1.601$\pm$0.05 & 0.036$\pm$0.004 & 0.647$\pm$0.019 \\
0.0,0.15 & 6.464$\pm$0.023 & 6.041$\pm$0.208 & 1.52$\pm$0.063 & 0.034$\pm$0.003 & 0.573$\pm$0.018 \\
0.2,0.15 & 6.297$\pm$0.031 & 0.559$\pm$1.054 & 1.943$\pm$0.09 & 0.033$\pm$0.004 & 0.463$\pm$0.02 \\
0.4,0.15 & 6.131$\pm$0.084 & 0.131$\pm$0.458 & 2.534$\pm$0.303 & 0.024$\pm$0.006 & 0.405$\pm$0.051 \\
-0.8,0.25 & 5.891$\pm$0.057 & 0.576$\pm$2.093 & 3.222$\pm$0.293 & 0.073$\pm$0.016 & 0.941$\pm$0.063 \\
-0.6,0.25 & 6.081$\pm$0.038 & 4.178$\pm$1.504 & 2.084$\pm$0.111 & 0.083$\pm$0.017 & 0.987$\pm$0.041 \\
-0.4,0.25 & 6.359$\pm$0.018 & 5.533$\pm$0.219 & 1.623$\pm$0.046 & 0.064$\pm$0.005 & 0.886$\pm$0.023 \\
-0.2,0.25 & 6.441$\pm$0.021 & 5.276$\pm$0.473 & 1.549$\pm$0.056 & 0.04$\pm$0.005 & 0.665$\pm$0.017 \\
0.0,0.25 & 6.059$\pm$0.041 & 0.694$\pm$0.219 & 2.216$\pm$0.134 & 0.039$\pm$0.006 & 0.575$\pm$0.037 \\
-0.8,0.35 & 5.953$\pm$0.032 & 0.693$\pm$0.226 & 2.766$\pm$0.158 & 0.061$\pm$0.008 & 0.95$\pm$0.039 \\
-0.6,0.35 & 6.076$\pm$0.021 & 1.099$\pm$0.629 & 1.974$\pm$0.061 & 0.097$\pm$0.008 & 1.116$\pm$0.039 \\
-0.4,0.35 & 6.082$\pm$0.025 & 0.724$\pm$1.248 & 1.951$\pm$0.059 & 0.082$\pm$0.01 & 0.932$\pm$0.034 \\
-0.2,0.35 & 6.15$\pm$0.053 & 1.015$\pm$0.886 & 3.134$\pm$0.328 & 0.031$\pm$0.007 & 0.579$\pm$0.032 \\ 
   \hline
    \end{tabular}
    \label{para-maps}
\end{table*}

\begin{figure*}
	\centering
	\includegraphics[width=18cm,viewport=0 40 1400 710,clip]{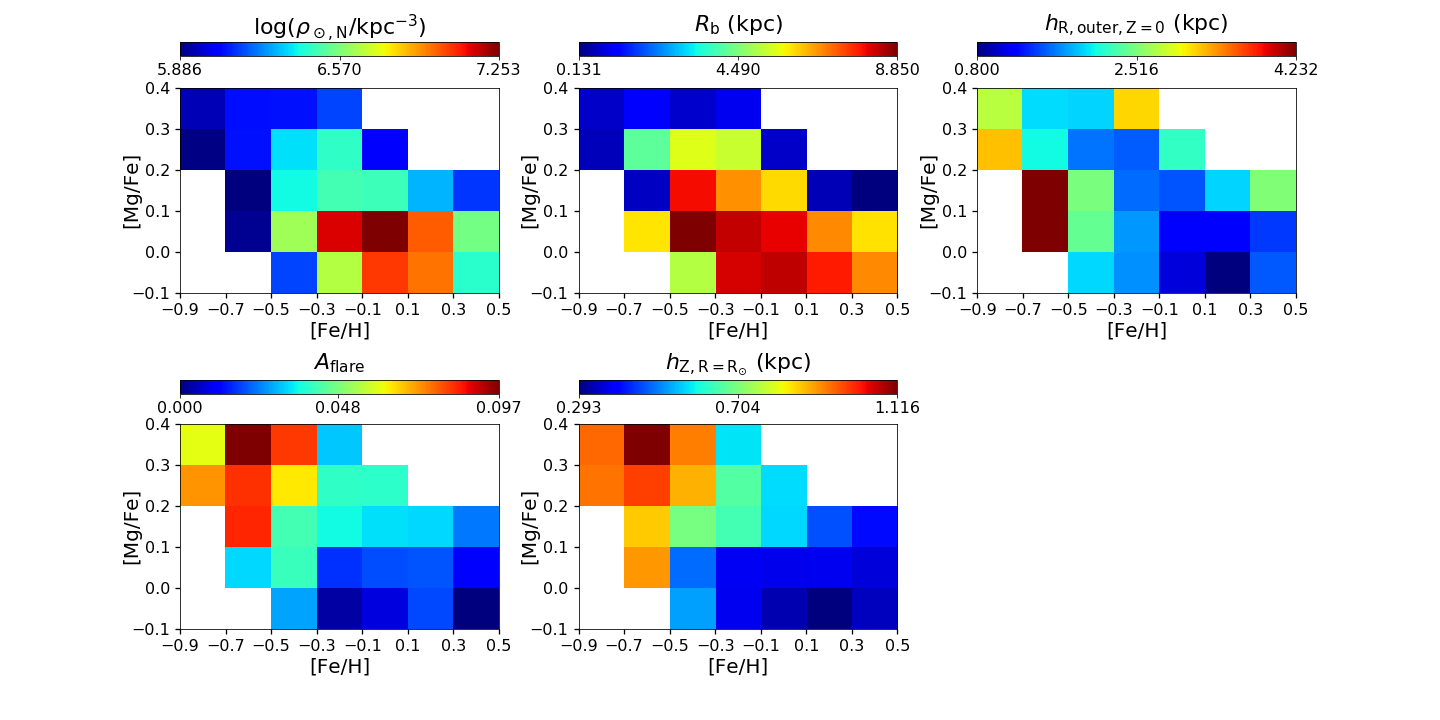}
	\caption{Distribution of best-fitted structure parameters in number density of mono-abundance populations in [Fe/H]-[Mg/Fe]. Each pixel indicates one mono-abundance population. Since a single exponential profile is used in the radial component of the density model for the high- and intermediate-$\alpha$ mono-abundance populations, the distribution of ${R_{\rm b}}$ and $h_{\rm R,inner}$ 
	in the middle and right top two panels are only populated by the low-$\alpha$ populations.} 
	\label{para-map-num}
\end{figure*} 

\begin{figure*}
	\centering
	\includegraphics[width=18cm,viewport=0 40 1400 710,clip]{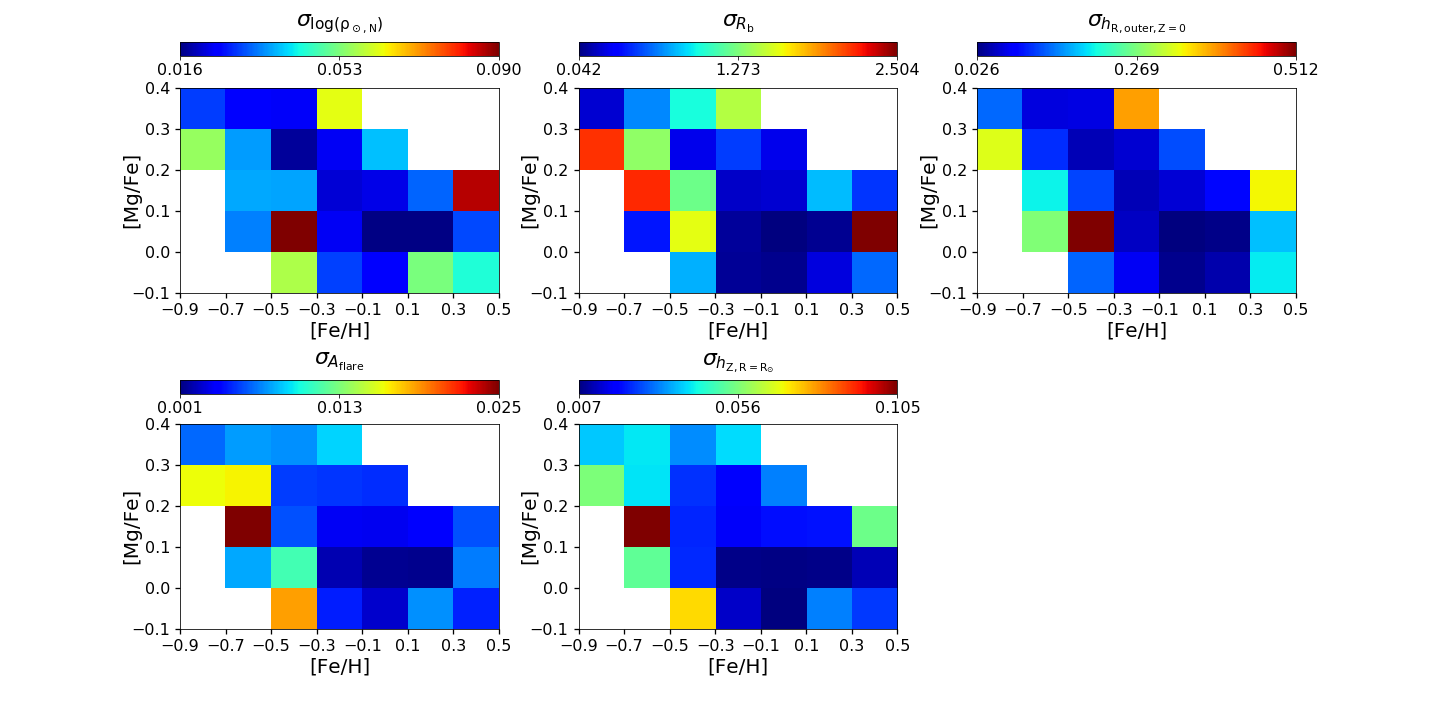}
	\caption{Uncertainties of best-fitted structural parameters for each estimated via Monte Carlo simulation. The uncertainty of derived intrinsic number density of the underlying population at each spatial bin is propagated from the uncertainty of the observed number density which is assumed to be Poisson error.} 
	\label{para-map-num-err}
\end{figure*} 
Figure~\ref{para-map-num} and Figure~\ref{para-map-num-err} present the distribution of the best-fitted parameters in [Fe/H]-[Mg/Fe] and their stochastic uncertainties estimated through Monte Carlo simulation. Poisson error is assumed for the observed number density at each spatial bin, which is then propagated to the derived intrinsic density of the underlying population. Each pixel corresponds to one MAP. The best-fitted parameters are also listed in Table~\ref{para-maps}.

$\rho_{\odot,N}:$ The local intrinsic number density of MAPs display large variations more than an order of magnitude with uncertainties less than 0.1~dex. The most common stars in the solar vicinity tend to have solar-like abundances, while the least numerous stars here are in the high-$\alpha$ sequence, with a decreasing number density at lower [Fe/H]. This selection-function corrected density distribution is qualitatively consistent with that seen in raw APOGEE data \citep[e.g.,][]{hayden2015}, confirming that APOGEE data have no significant selection bias on chemical abundances \citep{rojas2019}.   

$R_{\rm b}:$ The break radius of the broken radial profile is largest in low-$\alpha$ MAPs with sub-solar metallicities, smaller in the super-solar metallicity, low-$\alpha$ MAPs and the smallest in the high-$\alpha$ MAPs. Given the mean age of different MAPs, this indicates an age dependence of break radius, with a larger break radius in younger population, and reflects radial expansion of the Milky Way's disc throughout its history, also known as inside-out growth of the disc \citep[e.g.,][]{frankel2019}. The typical break radius of external galaxies with downbending broken profile (steeper outer region) is generally between 5-15~kpc \citep{pohlen2006}, comparable to the break radius we find here in the Milky Way.   

$h_{\rm R,inner}:$ The slope of the radial density distribution of MAPs in the inner Galaxy is consistently flat ($h_{\rm R,inner}$ generally greater than 100), in contrast to the rapid decrease beyond the break radius. Since $h_{\rm R,inner}$ presents no clear dependence on [Fe/H] and [Mg/Fe], this parameter is not included in Figs. 10-13. 

$h_{\rm R,outer}:$ The slope of the radial density distribution beyond the break radius presents a complicated distribution in abundance space. Interestingly, there is a similar trend of flatter radial density distribution at lower [Fe/H] in both high- and low-$\alpha$ MAPs. The origin of this trend is unclear, but not likely driven by radial migration. For high-$\alpha$ populations, they are generally old and have similar ages across metallicities. If they have experienced radial migration, the strength of migration should be similar. For low-$\alpha$ populations, their age does not monotonically correlate with the their [Fe/H] \citep[e.g.,][]{anders2017,feuillet2018}. In fact, the most metal-rich stars are actually the oldest population on average. However these metal-rich stars exhibit shorter outer scale length compared to low-$\alpha$ MAPs with [Fe/H]$<-0.1$, the opposite of what is expected if they have experienced more radial migration. 
The different outer scale length of the radial density distribution in the high-$\alpha$ and metal-rich, low-$\alpha$ populations suggest that they are not always physically associated at all radii of the Galaxy. This can be explained in disc formation models where the star formation and chemical evolution during the thick-to-thin disc transition is radially dependent (\citealt{chiappini2009,sharma2020}, Lian et al. in prep). Note that the shorter outer scale length in metal-rich, low-$\alpha$ populations than the high-$\alpha$ stars is not at odds with their similar radial extent as shown in Fig.~\ref{xy-rz}, since the low-$\alpha$ populations have a larger break radius. 

$A_{\rm flare}:$ This parameter characterizes the strength of flaring, which is defined here to be the slope of radial variation in scale height. The high-$\alpha$ MAPs generally flare, with the strongest flaring at [Fe/H]$=-0.6$ and decreasing strength at lower and higher [Fe/H]. This is different from the constant thickness of high-$\alpha$ populations reported in \citet{bovy2016}, but broadly consistent with the results in \citet{mackereth2017} and \citet{yu2021} where a flaring high-$\alpha$ disc was also found although with less strength than the low-$\alpha$ disc. 
In the low-$\alpha$ MAPs, the super-solar and solar-abundance stars present negligible flaring, while stars at lower [Fe/H] flare with moderate strength that is less than the high-$\alpha$ stars. 

$h_{\rm Z,R=R_{\odot}}:$ The scale height of MAPs at solar circle show a clear trend of decreasing thickness with lower [Mg/Fe]. 
Similar to the radial variation of scale height. The MAP with [Fe/H]$=-0.6$ and [Mg/Fe]$=0.35$ stands out with the widest vertical density distribution and a scale height of $\sim$1.1~kpc, while the solar-abundance stars are the most confined to the disc plane with scale height $\sim$0.3~kpc. 

\subsection{Structure parameters of mass and luminosity density distribution}
\begin{figure*}
	\centering
	\includegraphics[width=18cm,viewport=0 40 1400 710,clip]{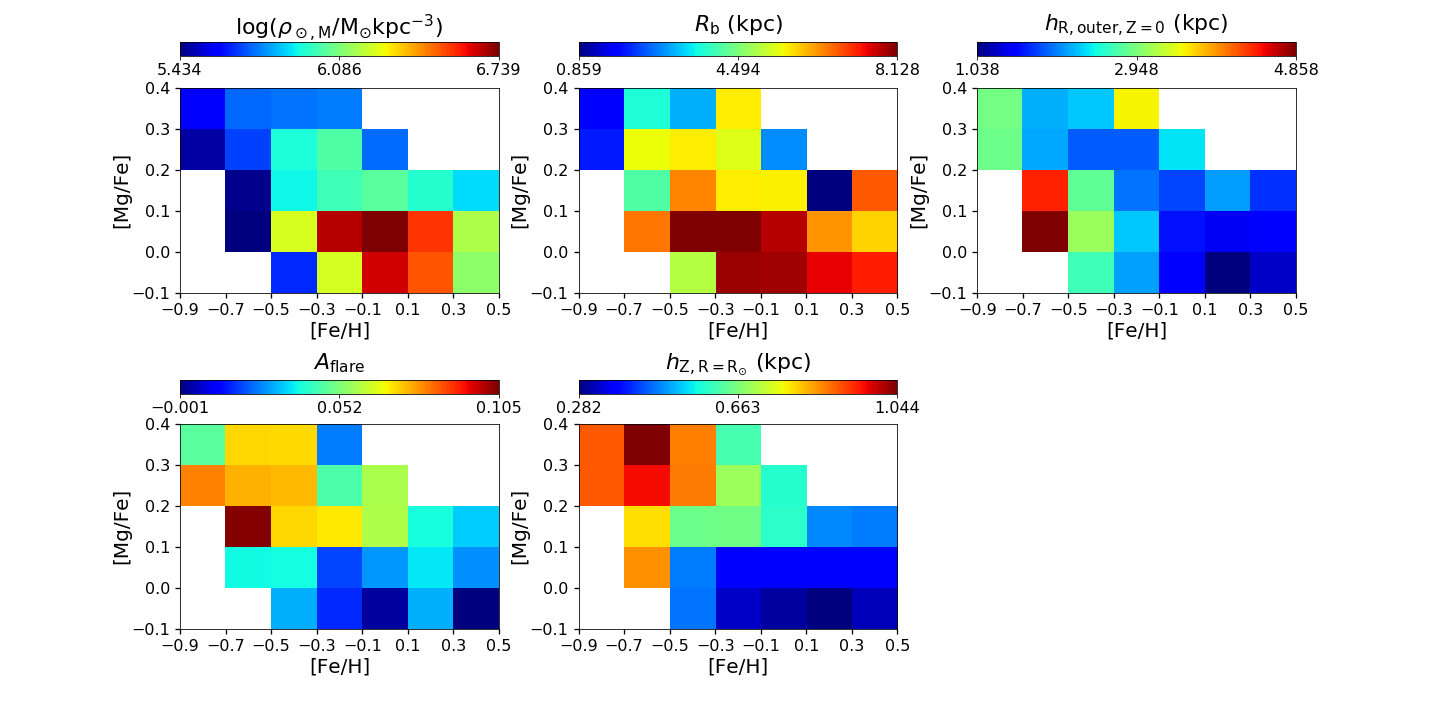}
	\caption{The same as Fig.~\ref{para-map-num} but the structure parameters are derived from the intrinsic mass density distribution.   
	} 
	\label{para-map-mass}
\end{figure*} 

\begin{figure*}
	\centering
	\includegraphics[width=18cm,viewport=0 40 1400 710,clip]{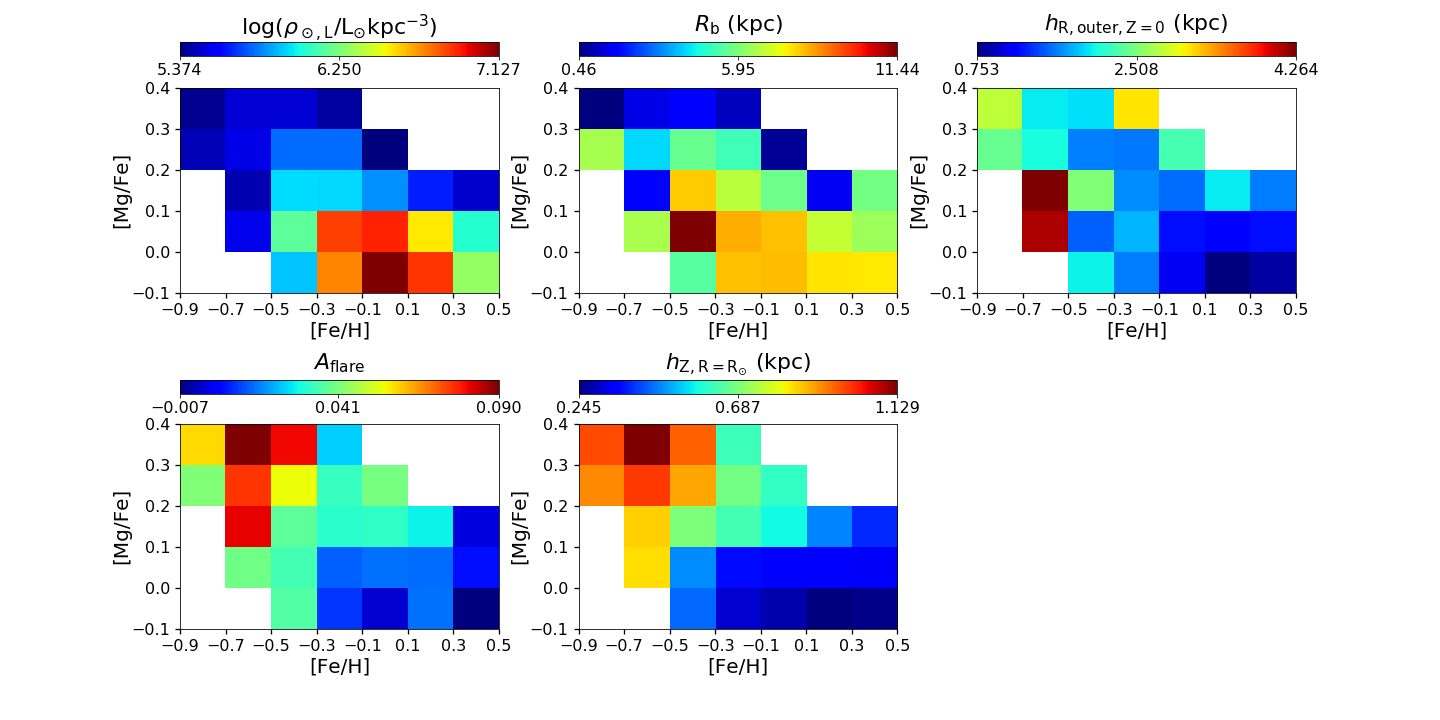}
	\caption{The same as Fig.~\ref{para-map-num} except that the structure parameters are derived from the intrinsic luminosity density distribution.   
	} 
	\label{para-map-lum}
\end{figure*} 

In addition to the intrinsic number density that is directly recovered from the data, it is also interesting to explore the spatial distribution of mass and luminosity of MAPs, which allows more direct comparison with the structure of Milky Way in the literature. 
Moreover, the obtained intrinsic mass and luminosity density distributions will lay the foundation to measure mass- and light-weighted integrated stellar population properties of our Milky Way as we do in other galaxies and enable direct comparison between them in the future. 

The conversion from number to mass and luminosity density is conducted by sampling the PARSEC isochrones \citep{bressan2012}, using the age distribution presented in \textsection2.2. For SDSS and 2MASS single band luminosities, PARSEC isochrones only provide absolute magnitude. 
To normalize to the solar luminosity, we use the solar magnitude in these filters \citep{willmer2018}.    
This sampling is performed for all locations where we have number density measurements.  
The obtained spatial distributions of mass and luminosity density have very similar shapes to those of the number density. 

We fit the spatial distributions of mass and bolometric/SDSS+2MASS luminosity densities of each MAP with the same density model and strategy as the fitting in number density. 
Figure~\ref{para-map-mass} and Figure~\ref{para-map-lum} present the distribution of best-fitted structure parameters for the mass and bolometric luminosity of all MAPS, respectively. The structure of MAPs in number, mass or luminosity 
are indistinguishable. 
One very minor change is the slight shift of the MAP with the highest density, from the one with [Fe/H]$=0$ and [Mg/Fe]$<$0.05 to the other MAP at the same [Fe/H] but [Mg/Fe]$=$-0.05. This is simply because of the younger ages in the latter MAP. 


\section{Discussion}

\subsection{Comparing to previous MAP studies}

In this section we discuss our best-fitted MAP structural parameters in comparison with previous observational  \citep{bovy2012b,bovy2016,mackereth2017,yu2021} and theoretical MAP studies \citep{bird2013,stinson2013,minchev2015}. The absolute scale and global shapes of radial and vertical density distributions are in good agreement with previous works using early APOGEE \citep{bovy2016,mackereth2017} or LAMOST data \citep{yu2021}. Many newly discovered features, such as broken radial density distribution of low-$\alpha$ MAPs and flaring high-$\alpha$ MAPs, are confirmed in this work. Nevertheless, there are notable differences from earlier works in some structural parameters. One of the most striking results in this work is the structure of high-$\alpha$ disc, which presents a broken radial density distribution similar to the low-$\alpha$ disc and the strongest flaring among all MAPs. This new result may provide valuable insights into the thick disc formation of the Milky Way. More detailed comparison on each aspect of MAPs' structure, from the local mass density to the radial and vertical structure, is given below. 

\subsubsection{Local mass density}
\citet{mackereth2017} derived the local surface mass density near the solar circle as a function of age and [Fe/H], and find the highest surface density in the youngest age, solar-[Fe/H] population. This is consistent with our result of highest local density at solar-[Fe/H] and solar-[Mg/Fe], which is
the MAP with the youngest average age (Fig.~\ref{age-dtr-r}). In the next section we will present our surface mass density measurements for integrated high- and low-$\alpha$ populations and perform a more quantitative comparison with the results in \citet{mackereth2017} and other non-MAP works. 

\subsubsection{Radial structure}
For the radial distribution of low-$\alpha$ MAPs, \citet{bovy2016} suggested a broken exponential profile provides a better fit to the raw density distribution of APOGEE low-$\alpha$ stars than a single exponential. A broken profile is also adopted in following MAP studies in \citet{mackereth2017} and \citet{yu2021}. In this work, by directly recovering the density distribution of underlying populations, we explicitly show the presence of such broken radial density distribution of low-$\alpha$ stars and also find a similar broken profile in high-$\alpha$ stars. 

\citet{bovy2016} and \citet{mackereth2017} found that the break radius in low-$\alpha$ MAPs is anti-correlated with [Fe/H] with a range of 6$-$12~kpc. This is not confirmed with LAMOST data in \citet{yu2021}, in which a roughly constant break radius was found.  These different results might be (partially) due to the different spatial coverage of the two surveys, in particular in the inner Galaxy. In this work we find the break radius peaks at $\sim$11~kpc in the low-$\alpha$ MAP at [Fe/H]$=-0.4$ and $[{\rm Mg/Fe}]=0.05$ and decreases towards higher [Fe/H] with $R_{\rm b}\sim 6$~kpc at [Fe/H]$=0.4$ and $[{\rm Mg/Fe}]=0.05$, in better agreement with \citet{bovy2016} and \citet{mackereth2017}. 

For the scale length past the break radius, \citet{bovy2016} obtained an $h_{\rm R,outer}$ of 2.3~kpc for the high-$\alpha$ stars using a single exponential profile and a range of 1.2$-$2.8~kpc for the low-$\alpha$ stars with longer scale length in lower [Fe/H] stars. \citet{yu2021} found a broken radial density profile of high-$\alpha$ disc and fit it with a double-exponential profile. They obtained an outer scale length of 2.0~kpc for the high-$\alpha$ disc and 1.0$-$2.9~kpc for the low-$\alpha$ disc with different [Fe/H]. In this work, we 
find a range of $h_{\rm R,outer}$ with the high-$\alpha$ MAPs, from 1.6~kpc at [Fe/H]=-0.4 and [Mg/Fe]=0.25 to 3.2~kpc at [Fe/H]=-0.8 and [Mg/Fe]=0.25. For the low-$\alpha$ MAPs, a shorter $h_{\rm R,outer}$ with a range of 0.8$-$2.5~kpc is obtained. We confirm the trend of longer scale length in more metal-rich MAPs.     

\subsubsection{Vertical structure}
The scale height of MAPs at the solar radius is generally found to be highest in the high-$\alpha$ MAPs ($\sim$0.8-1~kpc) and to decrease continuously in less $\alpha$-enhanced populations ($\sim$0.3~kpc) in both the Milky Way \citep{bovy2012b,bovy2016,mackereth2017} and simulated disc galaxies \citep{stinson2013}. This result is quantitatively confirmed in this work, and is shown to be valid for the measurements made in either number, stellar mass, or luminosity density. By dissecting the disc into mono-age and mono-[Fe/H] populations, \citet{mackereth2017} further found a dependence of local scale height on stellar age, with a scale height that decreases in younger populations. \citet{bird2013} performed detailed analysis of mono-age populations in a Milky Way-like simulated galaxy, and found a similar trend of shorter scale heights in younger populations. In this work we find the shortest local scale height in solar-like abundance ([Fe/H]$=0$, [Mg/Fe]$=-0.05$) MAPs, which have the youngest age among all MAPs, suggesting a qualitative dependence of scale height on age, consistent with that found in \citet{mackereth2017} and \citet{bird2013}.   

Regarding the radial dependence of scale height, \citet{bovy2016} found that the scale height of low-$\alpha$ MAPs increases with radius, with steeper increase in more metal-poor MAPs. The authors explored a variety of parameterizations for this radial variation and chose an exponential form to fit the raw density distribution in APOGEE. 
\citet{mackereth2017} adopted the same exponential form for their mono-age, mono-abundance populations and reported a trend of decreasing flaring strength in older low-$\alpha$ MAPs. 
Using LAMOST data, \citet{yu2021} recovered the density distribution of underlying MAPs and found the scale height of low-$\alpha$ MAPs change little within 10~kpc and then increase rapidly beyond this radius. 
Since a linear description of flaring is adopted in this work while an exponential form was used in previous MAP studies, direct quantitative comparison on the strength of flaring is difficult. 
Qualitatively, we find moderate flaring in metal-poor, low-$\alpha$ ([Fe/H]=-0.4 and [Mg/Fe]$<0.2$) MAPs and weak flaring in more metal-rich MAPs,  consistent with \citet{bovy2016}. Since the super-solar metallicity, low-$\alpha$ MAPs are older than the metal-poor, low-$\alpha$ MAPs on average, their weaker flaring does not support a secular origin of flaring which would result in a stronger flaring in older populations. 
A similar radial dependence of the scale height is also seen in some low-$\alpha$ MAPs, with constant scale height in the inner Galaxy and a linear increase beyond, in agreement with \citet{yu2021}. 


For high-$\alpha$ MAPs, \citet{bovy2016} found a constant scale height across the Galaxy, while \citet{mackereth2017} showed some radial variation of scale height but at a lower level compared to the low-$\alpha$ MAPs which is also seen in \citet{yu2021}. Among the high-$\alpha$ MAPs, \citet{mackereth2017} found an interesting trend of stronger flaring in older high-$\alpha$ MAPs. In contrast, in this work we find a much stronger signature of flaring in the high-$\alpha$ MAPs (peak $A_{\rm flare}\sim$0.09), which is even stronger than in the low-$\alpha$ MAPs (peak $A_{\rm flare}\sim$0.04). In particular, the high-$\alpha$ MAP at intermediate metallicity ([Fe/H]$=-0.6$) exhibits the strongest flaring, with the strength decreasing at both lower and higher metallicity. 

Among all MAPs, we find the lowest level of flaring (least radial variation of scale height) in the solar-abundance MAP, which is the youngest MAP, and the strongest flaring in intermediate-[Fe/H], high-$\alpha$ MAPs which are among the oldest MAPs. 
This result, however, is inconsistent with the trend presented in \citet{mackereth2017} that the youngest MAP presents the strongest flaring. These opposite results are surprising. The reason for this discrepancy is still unclear. It is worth noting that, in two simulated Milky Way-mass disc galaxies, \citet{minchev2015} found a trend of increasing flaring strength in older mono-age populations with the strongest flaring in the old thick disc, in good qualitative agreement with the result in this work. The flaring in these two simulated galaxies is a combined result of environment effect and secular evolution. Using a larger sample simulated galaxies, \citet{cruz2021} found a large variety of flaring strength in their geometric thick disc, with the flared thick discs generally consist of flared mono-age populations with large surface mass density. 


\subsection{Structure of the total stellar populations}
\begin{figure*}
	\centering
	\includegraphics[width=13cm,viewport=0 10 1000 850,clip]{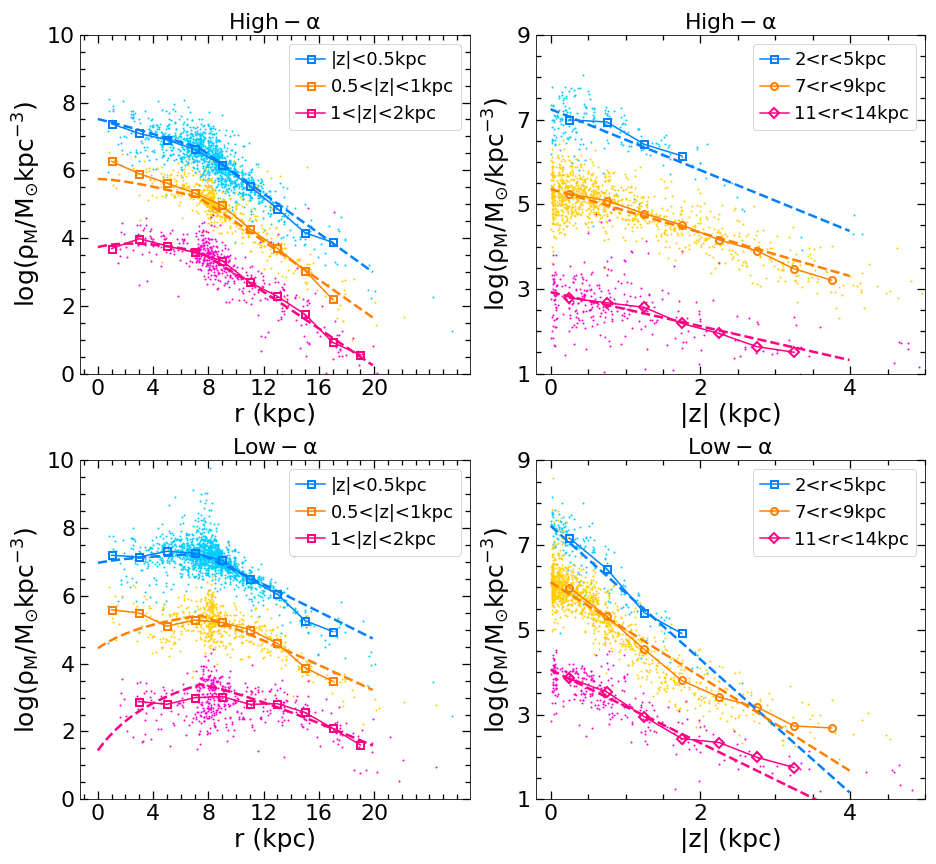}
	\caption{Left: Radial distribution of integrated intrinsic mass density of high- and low-$\alpha$ sequences (as defined in \textsection6.3) in the top and bottom panel, respectively. In each panel, the density measurements are separated into three wide $|{\rm Z}|$ height bins as shown in three different colours. Solid squares represent median mass density at each bin in radius and height. Right: Vertical distribution of integrated intrinsic mass density for the high-$\alpha$ sequence at the top and low-$\alpha$ sequence in the bottom. Density measurements are grouped into three wide radial bins denoted in different colours.    
	} 
	\label{den-disks}
\end{figure*}

\begin{figure*}
	\centering
	\includegraphics[width=15cm,viewport=0 0 900 400,clip]{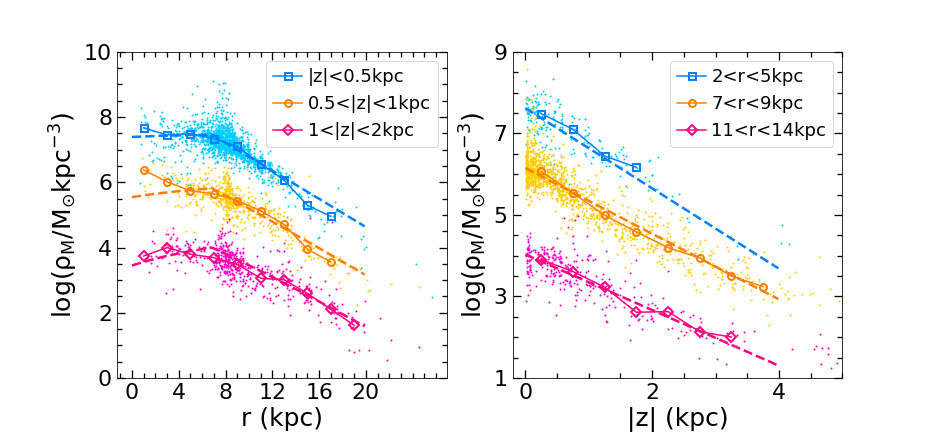}
	\caption{Similar to a row in Fig.~\ref{den-disks} but showing the spatial intrinsic mass density distribution of all mono-abundance populations together.     
	} 
	\label{density-whole}
\end{figure*}

In this section we derive the global structure of the Milky Way in mass and luminosity density distribution, which enables comparison with previous results where the stars are not finely split into sub-populations in abundance or age.   

\subsubsection{Integrated density distribution}

We first explore the density distribution of total populations on the high- and low-$\alpha$ sequence. 
Note that the high- and low-$\alpha$ MAPs refer to mono-abundance bins above and below [Mg/Fe]=0.2. To better separate the high- and low-$\alpha$ sequence and integrate the density for each sequence, we adopt the demarcation line from \citet{lian2020b} (shown in their Fig.~2), which is adapted from \citet{adibekyan2011} to APOGEE data. There are MAPs at $0.1<$[Mg/Fe]$<0.2$ and $-0.5<$[Fe/H]$<0.1$ that are divided by this line. For these three mono-abundance bins, we calculate the area in the [Fe/H]-[Mg/Fe] space within the high- and low-$\alpha$ regime and use that as an estimate of the fraction of each MAP in the two sequences. We then sum up the density at each location of MAPs in the high- and low-$\alpha$ sequences separately.
Figure~\ref{den-disks} presents the radial (left) and vertical (right) distribution of intrinsic mass density for the integrated high- (top) and low-$\alpha$ (bottom) population. Similar to Fig.~\ref{den-rprof} and Fig.~\ref{den-zprof} for individual MAPs, the radial (vertical) distribution are split into three vertical (radial) bins as illustrated in the top right legend. 

For the integrated high-$\alpha$ population, the vertical density profile can be well described by a single exponential component and the radial density profile tend to flatten within solar radius. 
The more significant flattening at $1<|Z|<2$~kpc in the inner Galaxy is largely due to strong flaring as discussed in \textsection4.2. 
For the integrated low-$\alpha$ population, the broken radial density profile is more predominant than the integrated high-$\alpha$ population, with a break near the solar radius. 
The radial density distribution at $0.5<|Z|<1$~kpc presents a more complicated shape with an excess at $R<4$~kpc, which is likely a signature of the bar.  
The vertical density distributions at intermediate and large radii show clear signatures of flattening at large vertical distances. This flattening feature is likely a result of different vertical structure within low-$\alpha$ MAPs; the steep and relatively flat profiles at $|{\rm Z}|<2$ and $|{\rm Z}|>2$~kpc are mainly contributed by the thin, metal-rich and thick, metal-poor low-$\alpha$ MAPs, respectively. 
 
In addition to the integrated high- and low-$\alpha$ populations, we also derive the intrinsic density of the total population by adding all considered MAPs ($-0.9<$[Fe/H]$<0.5$) together.  
Figure~\ref{density-whole} shows the radial (left) and vertical (right) mass density distribution of this total population. The shape of the radial density profile is more similar to the integrated low-$\alpha$ population with a clear break at solar radius as the low-$\alpha$ stars has a larger fraction of the total mass than the high-$\alpha$ stars. 

\subsubsection{Best-fitted structure parameters}
\begin{table*}
    \caption{Best-fitted structure parameters of integrated high-$\alpha$, low-$\alpha$, and total population weighted by stellar mass and three different luminosities.}
    \centering
    \begin{tabular}{l l c c c c c c }
    \hline\hline 
    & & $\rho_{\odot}$ & $R_{\rm b}$ & $h_{\rm R,outer}$ & $A_{\rm flare}$ & $h_{\rm Z,R=R_{\odot}}$ & Surface density\\
    & & log($\rm M_{\odot}/kpc^{3}$) & kpc & kpc & - & kpc & ${\rm M_{\odot}(L_{\odot})/pc^2}$ \\
    \hline
& Stellar mass & 6.52$\pm$0.008 & 7.309$\pm$0.081 & 1.467$\pm$0.012 & 0.057$\pm$0.0025 & 0.852$\pm$0.011 & 5.619$\pm$0.076 \\
High-$\alpha$ & Bolometric luminosity & 6.44$\pm$0.008 & 7.326$\pm$0.092 & 1.525$\pm$0.014 & 0.055$\pm$0.0026 & 0.854$\pm$0.01 & 4.719$\pm$0.075 \\
& r-band luminosity & 6.276$\pm$0.009 & 7.348$\pm$0.096 & 1.55$\pm$0.015 & 0.054$\pm$0.0024 & 0.865$\pm$0.01 & 3.259$\pm$0.051 \\
& H-band luminosity & 6.733$\pm$0.009 & 7.309$\pm$0.095 & 1.509$\pm$0.014 & 0.054$\pm$0.0026 & 0.834$\pm$0.011 & 9.026$\pm$0.143 \\
\hline
& Stellar mass & 7.3$\pm$0.01 & 7.456$\pm$0.098 & 2.075$\pm$0.024 & 0.027$\pm$0.0005 & 0.393$\pm$0.003 & 15.689$\pm$0.318 \\
Low-$\alpha$ & Bolometric luminosity & 7.502$\pm$0.01 & 7.905$\pm$0.101 & 1.999$\pm$0.028 & 0.026$\pm$0.0004 & 0.366$\pm$0.003 & 23.318$\pm$0.490 \\
& r-band luminosity & 7.348$\pm$0.01 & 8.047$\pm$0.097 & 2.016$\pm$0.028 & 0.026$\pm$0.0005 & 0.369$\pm$0.003 & 16.448$\pm$0.362 \\
& H-band luminosity & 7.752$\pm$0.009 & 7.848$\pm$0.062 & 1.962$\pm$0.024 & 0.026$\pm$0.0004 & 0.367$\pm$0.002 & 41.539$\pm$0.758 \\
\hline
& Stellar mass & 7.32$\pm$0.002 & 6.824$\pm$0.022 & 1.977$\pm$0.01 & 0.023$\pm$0.0004 & 0.542$\pm$0.002 & 22.666$\pm$0.108 \\
Total & Bolometric luminosity & 7.482$\pm$0.003 & 7.474$\pm$0.017 & 1.967$\pm$0.011 & 0.02$\pm$0.0003 & 0.493$\pm$0.001 & 29.906$\pm$0.180 \\
& r-band luminosity & 7.32$\pm$0.003 & 7.516$\pm$0.022 & 1.997$\pm$0.009 & 0.02$\pm$0.0003 & 0.496$\pm$0.001 & 20.761$\pm$0.137 \\
& H-band luminosity & 7.737$\pm$0.003 & 7.406$\pm$0.025 & 1.927$\pm$0.009 & 0.02$\pm$0.0003 & 0.493$\pm$0.001 & 53.898$\pm$0.362 \\
    \hline
    \end{tabular}
    \label{para-int}
\end{table*}

In order to obtain the structure parameters of the integrated populations, we perform a similar density fitting as we do for mono-abundance populations. 
We fit the spatial distribution of intrinsic mass and luminosity density of the integrated  populations with the same density model used for the MAPs i.e. a single exponential profile for the vertical density distribution and a double-exponential profile for the radial density distribution. 

The best-fitted structure parameters are included in Table~\ref{para-int}. 
To estimate uncertainties, we adopt a Monte Carlo simulation approach, which is also used for the fitting of MAPs in \textsection5.1. We first recompute the density distribution considering Poisson error in the raw number of APOGEE stars in each spatial bin, and then perform the fitting to each resampled distribution. 
After repeating this 100 times, we take the standard deviation of each fitted parameter as the uncertainty of that structural parameter. 

The uncertainties in the selection function are difficult to determine and are not considered in this work. Potential sources of uncertainty include the choice of isochrone, IMF, and possibly the assumption of intrinsic age distribution. Therefore the structural parameter uncertainties in Table~\ref{para-int}, which only include uncertainties from the density sampling and fitting, should be considered lower limits. 

With the best-fitted density model, we obtain the local surface mass density of each integrated population at the solar radius and compare them with literature results. 
The derived local surface mass density of high-$\alpha$, low-$\alpha$, and total stars are 5.62$\pm$0.0.08, 15.69$\pm$0.32, and 22.67$\pm$0.11 ${\rm M_{\odot}pc^{-2}}$, respectively. Based on mono-age-abundance population analysis with a forward modelling approach, \citet{mackereth2017} obtained a surface mass density of 3.0$^{0.4}_{-0.5}$, 17.1$^{2.0}_{-2.4}$, and $20^{+2.4}_{-2.9}$ for these three integrated populations, respectively. 
The measurement of the integrated low-$\alpha$ population of this work is consistent with \citet{mackereth2017}, while those of the high-$\alpha$ and total populations are higher in this work.  
The higher surface mass density of the high-$\alpha$ population in our work is mainly because we include those MAPs at [Fe/H]$<-0.5$, which comprise 33\% of the total high-$\alpha$ populations. If only considering high-$\alpha$ population at [Fe/H]$>-0.5$, the surface mass density becomes 3.76~${\rm M_{\odot}pc^{-2}}$, which agrees with the value reported in \citet{mackereth2017} within 2$\sigma$. 
Including these metal-poor high-$\alpha$ MAPs also results in the higher surface mass density of the total integrated populations in this work. 
Since the metal-poor high-$\alpha$ stars are included, we obtain a local density ratio between the (chemical) thick to thin disc of 17\% and local surface mass density ratio of 36\%, with the latter higher than \citet{mackereth2017} result (18\%). 

Before the advent of massive stellar spectroscopic surveys, many efforts have been devoted to study the structure of the Milky Way based on photometric datasets. 
Interestingly, a much less massive thick disk, compared to the findings here is usually reported in the literature with an average local thick-to-thin density ratio of 4\% and surface mass density ratio of 12\% according to the compilation of literature results in \citet{bland2016}. A more massive thick disk, however, has also been suggested in many works \citep[e.g.,][]{fuhrmann2008,juric2008,snaith2014}. For example, \citet{juric2008} obtained a local thick to thin disc density ratio of 12\%, in good agreement with our result. It is worth pointing out that a geometric definition of thick and thin disc is generally adopted in these studies. Since the geometric thick disc, near the solar circle, is dominant at $|{Z}|>1$~kpc and its contribution to the total mass density decreases to the midplane \citep[e.g.,][]{gilmore1983}, the vertical structure of the geometric thick disc relies heavily on the star counts at large Galactic latitude, which only comprise a minor fraction of the surface mass density of the thick disc. The significant vertical extension of low metallicity, low-$\alpha$ stars (see Fig.~\ref{xy-rz}) suggests a non-trivial contamination of these stars at large height. These two facts complicate the comparison between the structure of geometrically and chemically defined thick disc. 
Beyond the solar radius, the connection between the chemically and geometrically defined thick and thin disc is more complicated and still under debate \citep{martig2016-age-gradient}. 

Our total local surface mass density is higher than \citet{mackereth2017}, but both are lower than most of the literature results which is of order $\sim30{\rm M_{\odot}pc^{-2}}$ \citep{flynn2006,bovy2012a,mckee2015}. As discussed in \citet{mackereth2017}, the red giant stars targeted by APOGEE are only a tiny fraction of the total underlying population in number and mass density and the conversion between these two would be very sensitive to any potential systematics in APOGEE stellar parameters in addition to the isochrone set and IMF adopted. 

The scale lengths past the break radius measured  in stellar mass, bolometric and single-band luminosity density distribution are very similar. The high-$\alpha$ disc presents a systematically shorter scale length comparing to that of the low-$\alpha$ disc (1.5 versus 2~kpc). 
These measurements, however, are generally shorter than the findings in the literature. For the (geometric) thick disc, a variety of scale length is reported in the literature, with 
a range of 1.8 to 4.9~kpc in various studies \citep[e.g.,][]{larsen2003,bensby2011,cheng2012,bovy2015,yu2021} and an average value of 2.0$\pm$0.2~kpc \citep[see the review in][]{bland2016}.  
For the lower-$\alpha$ population, a scale length of $\sim$2.6~kpc is generally suggested in previous studies  \citealt{juric2008,bland2016}. Note that all these measurements in the literature are based on a fundamental assumption that the density distribution follows a single exponential profile. The shorter scale length in this work is mostly because we find that the radial mass and luminosity density distribution of both low- and high-$\alpha$ disc deviates from a single exponential profile and we adopt a double-exponential model to fit the data. If we change the configuration of the density model we get a different result. For example, if we use a single exponential model we arrive at a larger scale length of 2.2~kpc for the integrated low-$\alpha$ population. If we further restrict the fitting to the radial range from 4 to 14~kpc, we obtain a scale length of 2.7~kpc, in good agreement with other studies. However, we want to emphasize that a single exponential profile does not characterize the radial density distribution of the thick, thin, or the whole disc. Using the scale length of a single exponential profile to derive the disc properties on large scale, such as total stellar mass, or half-mass/light radius, would result in incorrect result. We will explore this point in a follow up work.     


The mass-weighted scale height of the integrated high- and low-$\alpha$ populations at the solar radius is 0.85 and 0.39~kpc, respectively. These results are consistent with that of high- and low-$\alpha$ MAPs presented above, and also in agreement with recent works that adopt a similar definition of high- and low-$\alpha$ populations using spectroscopic abundances \citep{bovy2016,yu2021}.  
\citet{juric2008} fit their Galactic density model with a large photometric sample of M dwarfs from SDSS and found scale heights of 0.9 and 0.3~kpc for the geometric thick and thin discs, respectively. 
The total population has a scale height of 0.78~kpc, in between that of the high- and low-$\alpha$ populations. 
It is interesting to note that the luminosity-weighted scale height of total integrated population is slightly shorter than the mass-weighted one, suggesting a slightly higher mass-to-light ratio above the disc. This is consistent with the slight positive age gradient in the vertical direction as shown in Fig.~\ref{age-dtr-z}.  

The integrated high-$\alpha$ population exhibits much stronger flaring compared to the integrated low-$\alpha$ population. This is again consistent with the trend we find in invidual MAPs but at odds with the constant scale height of high-$\alpha$ population reported in \citet{bovy2016}. \citet{yu2021} found a clear signature of flaring in high-$\alpha$ population, in line with this work, but the strength seems lower than the low-$\alpha$ population (by visual inspection of their Fig.~6).   
There is no significant difference in the structure parameters measured in optical or near-IR band compared to those in bolometric luminosity. This supports direct comparison between various works based on observations at different wavelengths.

\subsection{Implications for the Galactic disc formation}
\subsubsection{Inside-out disc formation}
An inside-out disc formation picture has been frequently invoked to interpret Galactic observations and Milky Way-like galaxy simulations \citep[e.g.,][]{larson1976,matteucci1989,bird2013,kobayashi2011}. 
The relatively compact radial extent of old, high-$\alpha$ MAPs compared to the young, solar-like abundance MAPs, as shown in Fig.~\ref{xy-rz}, suggests that recent star formation is more radially extended than the early star formation that built up the high-$\alpha$ disc. 
This seems to support an overall inside-out formation of the Milky Way's disc, with
the inner Galaxy built up rapidly with a compact morphology in the early times and the bulk of the outer Galaxy formed more recently. 
The relatively flat radial profile of total gas mass density in the Milky Way \citep{nakanishi2016} suggests that the extended star formation is likely due to recent gas accretion preferentially onto the outer Galaxy, which has also been suggested in many disc formation models to explain the radial variation of stellar distribution in [$\alpha$/Fe]-[Fe/H] \citep[e.g.,][]{chiappini2009,andrews2017,sharma2020,johnson2021,lian2020a,lian2020b}.  

Interestingly, as shown in Fig.~\ref{xy-rz}, the metal-poor, low-$\alpha$ MAP ([Fe/H]$=-0.4$ and [Mg/Fe]$=$0.15), which has intermediate age, exhibits the widest radial extension and longest scale length. \citet{bovy2016} and \citet{mackereth2017} also found the most broadened radial profile around the peak density for their most metal-poor MAP in the low-$\alpha$ sequence. \citet{mackereth2017} interpreted this broadened radial distribution as a result of radial migration that flattens the density distribution. The flat density profile of the low-$\alpha$ MAPs within the break radius found in this work may also be caused by radial migration. However, the youngest solar-abundance MAP, which should not have experienced much radial migration (Lian et al. in preparation), also shows this behavior, which implies a possibly different origin of the flat radial density distribution in the inner Galaxy. For example, interactions between Sagittarius and the disk may result in significant radial mixing \citep{carr2022}.


In addition to the inside-out formation, an `upside-down' formation picture of the disc is often proposed to explain the more extended vertical distribution of the old thick disc comparing to the relatively young thin disc \citep{bird2013,freudenburg2017}. 
A negative radial age gradient at large distance above the plane was identified by \citet{martig2016}, which complicates the picture of thick disc formation. 
Our results show that the metal-poor, low-$\alpha$ MAPs, which are geometrically thick but have ages a few Gyrs younger than the high-$\alpha$ MAPs, 
are more radially extended. It is then conceivable that the geometric thick disc at large radii is a superposition of old, high-$\alpha$ and intermediate-age, low-$\alpha$ populations, with increasing fraction of the latter at larger radii. This would naturally give rise to a negative age gradient above the disc plane. 
Therefore, there may be two episodes of thick disc formation in the Milky Way are separated in time and physical space, with the earlier phase more centrally concentrated and the later phase more radially extended. 

Among the high-$\alpha$ MAPs, we find a variety of radial and vertical structure, with the intermediate-[Fe/H] high-$\alpha$ MAPs being the least radially extended, the thickest and the most flared. This is not likely a result of inhomogeneous formation in early times, given the
nearly identical [$\alpha$/Fe]-[Fe/H] distribution of high-$\alpha$ sequence across the Galaxy \citep{weinberg2019,katz2021},
but rather suggests possible various structure evolution of high-$\alpha$ MAPs after thick disc formation. 
Since the high-$\alpha$ MAPs, especially the metal-poor ones ([Fe/H]$<-0.7$), are generally centrally concentrated, the observed number of high-$\alpha$ stars in the outer Galaxy is very low and the recovered intrinsic density distribution is subject to more stochastic uncertainties. The obtained absolute values of structure parameters of metal-poor, high-$\alpha$ MAPs should be used with caution, and we look forward to more robust measurements using datasets with better sampling in the outer Galaxy in the future. 

\subsubsection{Origin of disc flaring}
The origin of disc flaring is still unclear. Interactions with satellite galaxies or radial migration has been suggested as possible explanations in the literature \citep[e.g.,][]{minchev2015}. The findings in this work that high-$\alpha$ and metal-poor, low-$\alpha$ MAPs stand out among all MAPs with large scale height and clear flaring suggest intriguing connection between the formation of the two populations  possibly with a common origin of their flaring. The relatively weak flaring in metal-rich, low-$\alpha$ MAPs, which are formed in between the previous two populations, serves as evidence against secular evolution process as potential driver of the disc flaring. As enlightened in \citet{lian2020a}, a short period star burst event, possibly triggered by close interaction or merging with a satellite galaxy, might be the common origin of the formation and flaring of the high-$\alpha$ and metal-poor, low-$\alpha$ populations. This star burst event is suggested to be able to explain the locus of these two populations in [$\alpha$/Fe]-[Fe/H], in particular the enhanced $\alpha$ abundance (the metal-poor, low-$\alpha$ population is mildly $\alpha$-enhanced with [$\alpha$/Fe]$\sim$0.15). It is possible that when the star burst associated with a galaxy interaction/merging event occurs the star formation at larger radii would be more vertically extended in response of the tidal force due to lower stellar density. 
{In this scenario, the formation and flaring of the high-$\alpha$ and metal-poor, low-$\alpha$ discs are closely related to the merger history of the Milky Way. Early galaxy mergers (e.g., Gaia-Enceladus at $\sim$10~Gyr ago; \citealt{helmi2018}) may be responsible for the thick and flaring structure of the high-$\alpha$ disc, while more recent merger(s) (e.g., Sagittarius dwarf galaxy at several Gyr ago; \citealt{kruijssen2020}) lead to the formation of the flaring metal-poor, low-$\alpha$ disc.}
Note that a different implication on the origin of flaring is expected from \citet{mackereth2017} in which the youngest populations were found to have the strongest flaring. 


\subsection{Potential systematics in structure parameters}

The structure parameters of MAPs in this work are derived by fitting a density model to the intrinsic density distribution of underlying populations after correcting for the APOGEE selection function. 
The conversion of number density from the observed giant stars, which are the main targets of the APOGEE survey, to the underlying population including main-sequence stars is as large as 3-7 order of magnitudes, depending on the Galactic location, as shown in Fig.~\ref{select-final}. This conversion also magnifies potential systematics in the data, and itself depends on the choice of isochrone and extinction map.   
Here, we explore potential sources of systematic uncertainty, including small number statistics, distance measurement of APOGEE stars, and the choice of isochrones and extinction maps.  

APOGEE target selection is performed on a field basis. The final effective selection function that we derived is a function of individual fields. The number of observed stars in a typical disc field is around several hundred \citep{zasowski2017,beaton2021}. Since we have 24 mono-abundance and 13 distance bins (312 independent bins) for each field, this would give us 1-3 stars on average at each mono-abundance and distance if we computed selection functions on a field-by-field basis. Assuming a Poisson error, many of the mono-abundance and distance bins would have zero stars populated and can not be included to constrain the density model. In this case, the best-fitted density model are only constrained by the locations populated by APOGEE stars and will therefore overestimate the density distribution. The degree of overestimate depend mainly on the average number of observed stars in each spatial bin, which determines the fraction of locations with zero number of stars.  

For this reason we rebin the fields in Galactic longitude and latitude to achieve better statistics. There are 2.7 fields on average in each longitude and latitude bin. This largely reduces the fraction of spatial locations with zero APOGEE stars due to stochastic variation with low statistics. The adopted size of spatial rebinning is a compromise of acceptable star counts in each location and spatial resolution. To verify that our result is robust with the current statistic after spatial rebinning, we perform a test using a coarser spatial binning scheme that results in a combination of 5.2 fields at each spatial location on average. This coarser binning yields number density distribution of MAPs that have better statistics at each location but poorer spatial resolution. We then fit the new density distribution as we do before in \textsection5.1. The derived structure parameters of each MAP is shown in Figure~\ref{para-map-num-lows}. It can be seen that the structure parameters are in good agreement with the ones derived with the original spatial binning, both in absolute values and the trend with abundances. Therefore we conclude that our result is robust with the current spatial binning.  

Another possible source of systematics that may affect our results is the distance of APOGEE stars. To test whether our result is dependent on the particular distance catalog that is used, we also use the distances provided by the StarHorse Value Added Catalog{\interfootnotelinepenalty10000 \footnote{\url{https://www.sdss.org/dr16/data_access/value-added-catalogs/?vac_id=apogee-dr17-starhorse-distances,-extinctions,-and-stellar-parameters}}}
\citep{santiago2016,queiroz2018} in which the distances are determined using data from {\sl Gaia} in conjunction with APOGEE. 
Note that there are priors used for the StarHorse distance measurements, including priors for thin/thick discs, halo and bulge/bar. 
We repeat the analysis in \textsection5.1 to derive the structure parameters of MAPs, which is shown in Figure~\ref{para-map-num-sh}. Again we find structure parameters and their dependence on abundance consistent with our previous results using AstroNN distance.  

Other potential sources of systematics include the isochrones and extinction which are used to estimate the first component of APOGEE selection function, $f_{\rm CMD}$, the fraction of the underlying population satisfying the APOGEE candidate limits defined in ($H, J-K_0$) colour-magnitude space as discussed in \textsection3.1. To qualitatively understand whether and how our results are affected by the choice of isochrones and extinction maps, we performed two tests to calculate the APOGEE selection function and structure paramters of MAPs, one using MIST isochrones \citep{paxton2011,choi2016,dotter2016} to replace the PARSEC isochrones adopted previously and the other considering the VVV extinction map \citep{schultheis2014} in the bulge fields. The structural parameters of these two tests are consistent with the fiducial results presented in Fig.10; differences are generally less than 10\%. The flattening radial density distribution of MAPs in the inner Galaxy and strong flaring of high-$\alpha$ MAPs are also present. 
Therefore we conclude that our results are robust against the specific choice of isochrones and extinction maps.  

\section{Summary}
In this work we recover the spatial intrinsic density distribution of MAPs traced by stars observed by APOGEE, and derive their structure parameters. A careful treatment of APOGEE's selection function is taken into account to convert the observed number density of APOGEE stars to the density of their underlying population. The derived conversion factor ranges from 10$^3$ to 10$^7$. 

Prior to fitting a parametric model to the density distributions, we draw some conclusions from the selection-corrected density measurements.
The recovered intrinsic spatial density distributions vary significantly between different MAPs. The high-$\alpha$ MAPs exhibit the most compact radial extension but the widest vertical distribution, while the metal-poor, low-$\alpha$ MAPs are most extended in the radial direction, and the solar-abundance MAPs have the shortest vertical expansion. By studying the density distribution of underlying populations directly, we explicitly uncover a broken radial density profile of high- and low-$\alpha$ MAPs, which has been adopted in the density model in many previous works \citep{bovy2016,mackereth2017}. This broken profile, with a nearly flat distribution within the break radius, suggests a single exponential form is no longer a good description of the disc in the Milky Way. The origin of this flattening in the inner Galaxy is a new open question. The presence of bulge/bar and/or radial migration might play a role.   

For the parametric fitting, to examine the radial variation of MAPs' scale height, we first perform an intermediate fitting to their vertical density distribution at a series of narrow radial bins using a single exponential profile. Surprisingly, we find a steep positive correlation between the scale height of high-$\alpha$ MAPs and radius, suggesting these old, $\alpha$-enhanced populations strongly flare. The flaring is weaker in MAPs of lower [Mg/Fe], which are generally younger. This is qualitatively with the structure of mono-age populations in simulated galaxies in \citet{bird2013} and \citet{minchev2015}. We note that some low-$\alpha$ MAPs seem to show an interesting broken correlation between their scale height and radius, with negligible radial dependence in the inner Galaxy and increase beyond the break radius. 

We then conduct a full fitting to the 2D spatial distribution of number, mass, bolometric and single-band luminosity densities as a function of radius and height to derive the structure parameters of each MAP. A broken radial profile and a linear radial variation of scale height is adopted for all MAPs. 
A careful quality inspection on projected 1D radial and vertical distribution is performed to make sure that the density model indeed well describes the observed density distribution. 
The best-fitted structure parameters for the individual MAPs are presented in Fig.10-13. The strong radial variation of scale height in high-$\alpha$ MAPs is confirmed. We also find interesting internal variation of structure within high-$\alpha$ MAPs, with the shortest scale length, longest scale height at solar radius, and strongest flaring in the high-$\alpha$ MAPs with [Fe/H]$\sim-0.6$. If confirmed, this might suggest differential structure evolution after the establishment of the chemical thick disc. 

The spatial structure of low-$\alpha$ MAPs also vary with metal content.
Metal-poor MAPs have the longest scale length and mildly flares, while the metal-rich MAPs present shorter scale length and weaker flaring. 
The young, solar-abundance MAPs have the shortest scale height and exhibit negligible signature of flaring, suggesting the recent star formation in the Milky Way are strictly confined to the disc plane across the Galaxy.

In addition to individual MAPs, we also derive the density distribution and structure parameters of integrated high-$\alpha$, low-$\alpha$ and total populations. The integrated total population shows a similar density distribution to the low-$\alpha$ MAPs, with a complicated radial density profile. We estimate a surface mass density of 5.62$\pm$0.0.08, 15.69$\pm$0.32, and 22.67$\pm$0.11 ${\rm M_{\odot}pc^{-2}}$ for the integrated high-$\alpha$, low-$\alpha$, and total populations, respectively. These are in good agreement with \citet{mackereth2017} after considering the extra high-$\alpha$ MAPs at [Fe/H]$<-0.5$ considered in this work, but are lower than other literature values \citep[e.g.,][]{mckee2015}. Our measurements suggest a more massive thick disc than many previous works, with a local mass density ratio between (chemical) thick to thin disc of 17\% and a local surface mass density ratio of 36\%. The structure parameters of integrated low- and high-$\alpha$ populations are generally consistent with that of individual MAPs in each group. 

The structure of MAPs provide strong constraints on the models of Galaxy formation and evolution. While many features in the density distribution of MAPs are yet to be understood, more subtle features not captured by the smooth density model may be due to interesting substructures. 
It will also be interesting to explore the integrated population properties of the Milky Way and directly compare them with external galaxies to address the fundamental question of whether and how our Milky Way is a special galaxy from a stellar population perspective. 

\section*{Acknowledgements}
We are grateful to the referee for their useful comments that improve the clarity of the paper. This material is based upon work supported by the National Science Foundation under Grant No. 2009993. JAH and JI acknowledge support from National Science Foundation under Grant No. AST-1909897. J.G.F-T gratefully acknowledges the grant support provided by Proyecto Fondecyt Iniciaci\'on No. 11220340, and also from ANID Concurso de Fomento a la Vinculaci\'on Internacional para Instituciones de Investigaci\'on Regionales (Modalidad corta duraci\'on) Proyecto No. FOVI210020, and from the Joint Committee ESO-Government of Chile 2021 (ORP 023/2021).
We thank the E-Science and Supercomputing Group at Leibniz Institute for Astrophysics Potsdam (AIP) for their support with running the StarHorse code on AIP cluster resources.

Funding for the Sloan Digital Sky Survey IV has been provided by the Alfred P. Sloan Foundation, the U.S. Department of Energy Office of Science, and the Participating Institutions. SDSS-IV acknowledges
support and resources from the Center for High-Performance Computing at
the University of Utah. The SDSS web site is www.sdss.org.

SDSS-IV is managed by the Astrophysical Research Consortium for the 
Participating Institutions of the SDSS Collaboration including the 
Brazilian Participation Group, the Carnegie Institution for Science, 
Carnegie Mellon University, the Chilean Participation Group, the French Participation Group, Harvard-Smithsonian Center for Astrophysics, 
Instituto de Astrof\'isica de Canarias, The Johns Hopkins University, Kavli Institute for the Physics and Mathematics of the Universe (IPMU) / 
University of Tokyo, the Korean Participation Group, Lawrence Berkeley National Laboratory, 
Leibniz Institut f\"ur Astrophysik Potsdam (AIP),  
Max-Planck-Institut f\"ur Astronomie (MPIA Heidelberg), 
Max-Planck-Institut f\"ur Astrophysik (MPA Garching), 
Max-Planck-Institut f\"ur Extraterrestrische Physik (MPE), 
National Astronomical Observatories of China, New Mexico State University, 
New York University, University of Notre Dame, 
Observat\'ario Nacional / MCTI, The Ohio State University, 
Pennsylvania State University, Shanghai Astronomical Observatory, 
United Kingdom Participation Group,
Universidad Nacional Aut\'onoma de M\'exico, University of Arizona, 
University of Colorado Boulder, University of Oxford, University of Portsmouth, 
University of Utah, University of Virginia, University of Washington, University of Wisconsin, 
Vanderbilt University, and Yale University.

\section*{Data Availability}
The data underlying this article is from an internal incremental release of the SDSS-IV/APOGEE survey, following the SDSS-IV public Data Release 16 (using reduction pipeline version r13). This incremental catalog is anticipated to be made public in a future post-DR17 release. 

\bibliographystyle{mnras}
\bibliography{Jianhui}{}

\appendix 
\section{Testing possible systematics in the results}
\begin{figure*}
	\centering
	\includegraphics[width=18cm]{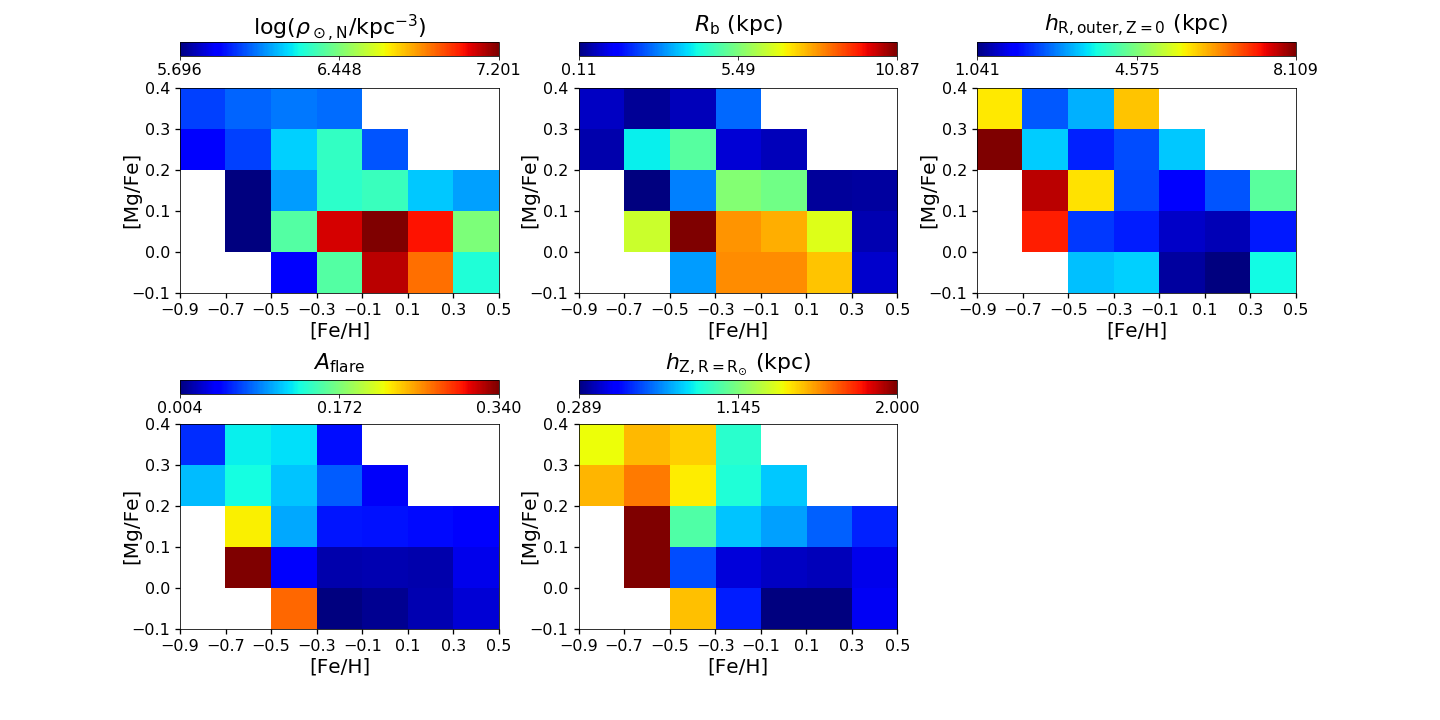}
	\caption{Distribution of best-fitted structure parameters in number density of [Fe/H]-[Mg/Fe] mono-abundance populations using spatial binning coarser than that adopted in the main paper.} 
	\label{para-map-num-lows}
\end{figure*}

\begin{figure*}
	\centering
	\includegraphics[width=18cm]{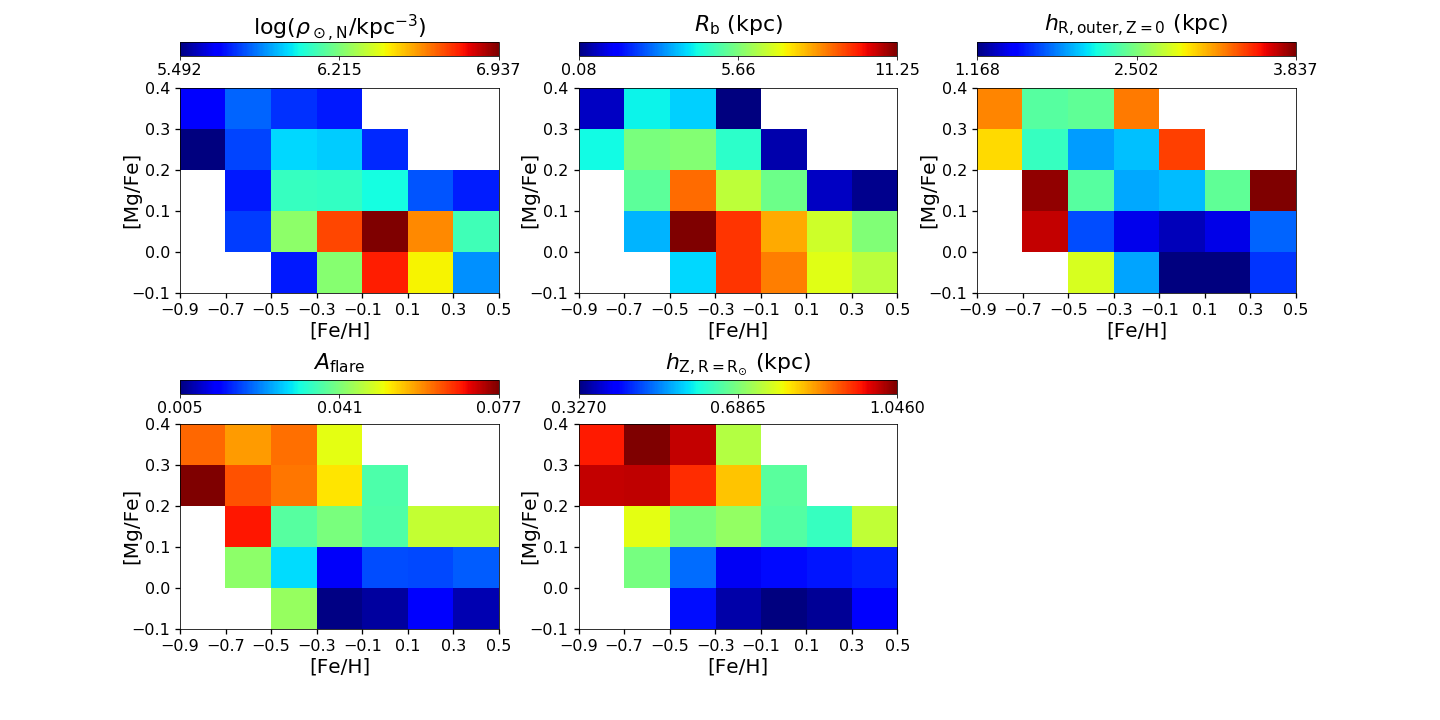}
	\caption{The same as Fig.~\ref{para-map-num} but using distances from StarHorse Value Added Catalog.} 
	\label{para-map-num-sh}
\end{figure*}

Here we present the test results of using a coarser APOGEE field binning scheme to examine the effect of low number star counts in Figure~\ref{para-map-num-lows} and stellar distances from StarHorse Value Add Catalog in Figure~\ref{para-map-num-sh}. The derived structure parameters and their dependence on abundances are consistent with those presented in the paper, suggesting that our results are robust against low number statistics and choice of stellar distance catalog. 

\end{document}